\documentclass[%
 amsmath,amssymb,
 aps,
 prd,
]{revtex4-2}

\usepackage{graphicx}
\usepackage{dcolumn}
\usepackage{bm}

\usepackage{graphicx}
\usepackage{latexsym}
\usepackage{bbold}
\usepackage{physics}
\usepackage{amssymb}
\usepackage{hyperref}
\usepackage{color}
\usepackage{empheq}

\def\be{\begin{equation}}
\def\te{\end{equation}}
\def\bea{\begin{equation}\begin{aligned}}
\def\tea{\end{aligned}\end{equation}}
\def\lc{\left(}
\def\rc{\right)}
\def\nn{\nonumber\\}

\begin{document}

\preprint{APS/123-QED}

\title{Hydrodynamic models of Reheating}

\author{Juan Pablo Elía}
\email{jp.elia@df.uba.ar}
\affiliation{Universidad de Buenos Aires, Facultad de Ciencias Exactas y Naturales, Departamento de Física, Buenos Aires, Argentina}
\affiliation{CONICET-Universidad de Buenos Aires, Instituto de Física de Buenos Aires (IFIBA), Buenos Aires, Argentina}

\author{Lucas Cantarutti}
\email{lcantarutti@df.uba.ar}
\affiliation{Universidad de Buenos Aires, Facultad de Ciencias Exactas y Naturales, Departamento de Física, Buenos Aires, Argentina}
\affiliation{CONICET-Universidad de Buenos Aires, Instituto de Física de Buenos Aires (IFIBA), Buenos Aires, Argentina}

\author{Nahuel Mirón-Granese}
\email{nahuelmg@df.uba.ar}
\affiliation{Consejo Nacional de Investigaciones Científicas y Técnicas (CONICET),
Godoy Cruz 2290, Ciudad de Buenos Aires C1425FQB, Argentina}
\affiliation{Universidad Nacional de La Plata, Facultad de Ciencias Astronómicas y Geofísicas, Paseo del Bosque, La Plata B1900FWA, Buenos Aires, Argentina }
\affiliation{Universidad de Buenos Aires, Facultad de Ciencias Exactas y Naturales, Departamento de Física,\\ Intendente Güiraldes 2160, Ciudad de Buenos Aires C1428EGA, Argentina}

\author{Esteban Calzetta}
\email{calzetta@df.uba.ar}
\affiliation{Universidad de Buenos Aires, Facultad de Ciencias Exactas y Naturales, Departamento de Física, Buenos Aires, Argentina}
\affiliation{CONICET-Universidad de Buenos Aires, Instituto de Física de Buenos Aires (IFIBA), Buenos Aires, Argentina}





\date{\today}

\begin{abstract}
We develop a causal hydrodynamic model that provides an effective macroscopic description of the field-theoretic dynamics during the early stages of reheating. The inflaton condensate is treated as a homogeneous background coupled to a relativistic fluid that represents its inhomogeneous fluctuations. Within the divergence-type theory framework derived from kinetic considerations, the model captures essential dissipative and non-equilibrium effects while remaining stable and causal. We find that the coupling between the oscillating condensate and the fluid induces a parametric resonance in the tensor sector, leading to the amplification of the viscous stress tensor and the generation of gravitational waves with a characteristic spectral peak. The predicted spectrum agrees with lattice simulations performed with CosmoLattice. This hydrodynamic approach offers an effective bridge between microscopic field dynamics and macroscopic cosmological observables.
\end{abstract}

\maketitle


\section{Introduction}

The era of reheating after inflation is now recognized as a distinct stage in the cosmological history of the Universe, alongside inflation itself and the radiation-, matter-, and dark energy–dominated eras~\cite{2019_Lozanov}. During reheating, the energy that powered the accelerated expansion is transferred into a hot plasma, thereby setting the initial conditions for the subsequent thermal history of the Universe.

Despite its central role, our knowledge of reheating remains limited. Most of the relevant dynamics during this process unfolds on subhorizon scales, whereas our most powerful observational probe of the early Universe, the Cosmic Microwave Background (CMB), is primarily sensitive to superhorizon physics at that time. The effect of reheating on the CMB is effectively encoded in a single parameter that characterizes the evolution of the scale factor during this epoch~\cite{2010_Martin,2015_Martin,2015_Cook,2024_Martin}. This parameter is sensitive to the equation of state during reheating and the final energy density, but not to the details of the reheating process itself. 

For this reason, the most promising avenues for probing reheating lie in relics formed at small scales and preserved until today \cite{2019_Lozanov}. Among these, a stochastic background of gravitational waves (GW) stands out \cite{1997_Khlebnikov,2006_Easther,2007_Easther,2008_Bellido}. Once produced, GW interact only weakly with matter, though not negligibly \cite{nahuel_esteban_2018,2021_Miron-Granese,baym_2017,nahuel_claudia_2025}, and therefore preserve a relatively clean imprint of their origin.

From the theoretical side, modeling reheating is challenging due to the wide hierarchy of scales involved. Much of our current understanding comes from field-theoretic models~\cite{2019_Lozanov,2006_Bassett,2010_Allahverdi,2014_Amin,2024_Mishra,2024_Kolb}. In these setups, the inflaton couples to a set of lighter fields that undergo parametric resonance and become exponentially amplified by the inflaton oscillations~\cite{1990_Dolgov,1990_Traschen,1994_KLS,1995_Shtanov,Boyanovsky_1996,1997_KLS}. In these scenarios interactions among the produced fields are often neglected or treated perturbatively. The relevant approximations break down once the inflaton condensate fragments~\cite{1996_Khlebnikov,1997_Khlebnikov_Tkachev,1997_Ramsey,2006_Podolsky,2023_Garcia,Garcia_mambrini_2023}. These models provide little insight into the subsequent thermalization of the highly excited matter fields~\cite{2002_Grana,2003_Micha,2004_Micha,2004_Berges,2020_McDonough}, which is essential for predicting the state of the Universe at the transition to radiation domination.

An alternative is the numerical simulation of the full nonlinear dynamics~\cite{2000_Felder,2008_Frolov,2023_Figueroa}. These simulations typically include the inflaton, a number of coupled matter fields, and the expanding Friedmann–Robertson–Walker geometry. Initial conditions are set by linearized quantum field theory at the end of inflation, and the system is evolved classically thereafter~\cite{2024_Figueroa}. Numerical studies have demonstrated that GW are indeed produced during reheating, mainly at early times, and have provided concrete predictions for their spectrum~\cite{2017_Figueroa,2018_Caprini,2025_Roshan}.

Nevertheless, numerical approaches face limitations similar to those encountered in turbulence simulations \cite{berselli2006mathematics,sagaut2006large}. Discretization of the degrees of freedom in either physical or momentum space inevitably introduces a tradeoff between the simulation size and the relevance of subgrid dynamics. Additional challenges arise from the large number of matter fields expected in realistic models, our ignorance of the detailed particle content of the dark sector, and the neglect of full quantum and general relativistic effects.

These challenges motivate the development of theoretical descriptions of reheating that go beyond early field-theoretic treatments, with the goal of aiding the interpretation of numerical results and, eventually providing effective subgrid models. A natural framework for such an approach is hydrodynamics.

Traditionally, hydrodynamics was regarded as a framework valid only once local thermal equilibrium is established, typically at relatively long time scales \cite{chapman_cowling}. However, studies of relativistic heavy-ion collisions have shown that hydrodynamics can successfully describe the evolution of the fireball immediately after the collision, at times far too early for equilibrium to be reached. This has motivated a revised understanding in which hydrodynamic behavior emerges well before thermalization, acting as an \emph{attractor} of the more complex underlying microscopic dynamics~\cite{STRICKLAND_2015,Heller_2015,Florkowski_2018,Romatschke_Romatschke_2019,heller_2022,Soloviev_2022}. Within this picture, one distinguishes between \emph{hydrodynamization}, the early onset of hydrodynamic behavior, and \emph{thermalization}, the later approach to local equilibrium. In this work, we regard hydrodynamics as the simplest effective theory consistent with conservation laws and the Second Law of Thermodynamics, while remaining both causal and stable.

Hydrodynamic models have previously been employed to gain qualitative insight into the turbulent cascade that redistributes the energy of parametrically amplified matter-field modes across the spectrum at early times~\cite{2002_Grana,2003_Micha,2004_Micha}. We aim to go beyond those approaches by developing more detailed and accurate hydrodynamic descriptions.

Our specific goal in this contribution is to construct a hydrodynamic model for the emission of gravitational waves during the early stages of reheating due to matter fluctuations. To this end, we adopt a simplified framework in which the inflaton is treated as a classical, homogeneous condensate, while all other degrees of freedom, including the inhomogeneous inflaton fluctuations, are collectively described as a “fluid”. At the times relevant for gravitational radiation, this fluid is in fact dominated by the inhomogeneous inflaton fluctuations.

As we shall show below, consistency requirements at the macroscopic level alone are not enough to single out an unique hydrodynamical model. To fill the gaps, we shall assume hydrodynamics is obtained from a more fundamental description, where the fluid excitations may be described as quasiparticles with a definite mass, obeying a kinetic equation of motion \cite{calzetta1988,jeon1995,jeon1996,calzetta2000,2008_Calzetta_Hu}. For tractability, we employ the relaxation-time approximation for the collision integral in kinetic theory \cite{bhatnagar_1954,chapman_cowling}.

The hydrodynamics of relativistic real fluids requires additional dynamical variables beyond the temperature, chemical potential, and velocity fields that define an ideal fluid. These extra degrees of freedom, appearing as short-lived, nonhydrodynamic modes in the linearized theory \cite{2021_Perna_Calzetta}, can either be absorbed into a choice of hydrodynamic “frame” (as in the Landau–Lifshitz or Eckart prescriptions) or included explicitly as new fields \cite{MITRA2025100054}. We adopt the latter approach.

Concretely, after writing down the kinetic equation for the quasiparticles interacting with the inflaton condensate, and the Klein–Gordon equation for the condensate, renormalized by the fluid backreaction, we postulate a parametrized form for the one-particle distribution function. The associated parameters define the degrees of freedom of the hydrodynamic model, and their dynamics follow from the moments of the kinetic equation \cite{2020_Cantarutti_Calzetta}, in a way that enforces the Second Law of Thermodynamics. These preliminary steps, necessary to obtain the hydrodynamic model, are detailed in Appendix \ref{Section_Derivation of Hydro equations}.

In the present model, the degrees of freedom include the usual temperature and velocity (with vanishing chemical potential, as appropriate in the absence of a matter–antimatter asymmetry) as well as two additional tensor fields. The first is a second-rank tensor associated with the viscous part of the energy–momentum tensor. The second is a third-rank tensor, required to allow tensor fluctuations to propagate with a definite speed~\cite{jaiswal2013,brito2022,2021_Perna_Calzetta,2022_Calzetta,brito2023,2024_deBrito,panday2024}.

Since the second-rank tensor is solely responsible for gravitational wave emission, in this first approach we neglect the inhomogeneous scalar and vector degrees of freedom of the fluid. Our model therefore consists of three components: a homogeneous, time-dependent inflaton condensate, and a fluid at rest with a time dependent, homogeneous temperatura and inhomogeneous tensor fluctuations of second and third rank. The third-rank tensor is subsequently eliminated, yielding a wave equation for the second-rank tensor. This tensor is then mapped onto the viscous part of the fluid energy-momentum tensor, which acts as the source of gravitational waves in the Einstein equations. The initial conditions for the inhomogeneous tensor field are obtained from the statistical fluctuations of a fluid in equilibrium at the end of inflation.

Our findings show that oscillations in the condensate induce a parametric amplification of the viscous energy-momentum tensor, whose decay into GW  results in a definite peak in the GW spectrum. The bandwidth and early-time growth of this peak are in good agreement with numerical simulations performed using the CosmoLattice code~\cite{2020_Figueroa,2023_Figueroa}, highlighting the potential of hydrodynamic models to provide phenomenological partners to the more detailed numerical descriptions. 

Several open issues remain for future work. These include a more realistic distinction between inflaton and matter, a complete treatment of all inhomogeneous fluid modes, the introduction of a mass spectrum in the fluid, the inclusion of long-range gauge fields and gravitational wave backreaction, a more realistic kinetic description, and a proper treatment of thermalization. Although a fully realistic hydrodynamic model will ultimately require numerical methods, solving the hydrodynamic equations remains considerably simpler than addressing the full field theory: even turbulent flows, while challenging, are more tractable than integrating the complete microscopic dynamics.

The rest of the paper is organized as follows. In the next section, we define the hydrodynamical model that effectively describes the Reheating period (Section \ref{Section_The model}). In Section \ref{Section_background}, we study the background dynamics, characterized by the fluid temperature and the condensate evolution. Section \ref{Section_Tensor perturbation dynamics} presents the equations for the tensor perturbations of the fluid, which are the only perturbations considered in this work. Section \ref{Section_Spectras} is devoted to determining the initial fluctuations of the fluid and defining the energy density spectra of both the fluid and the gravitational waves produced. In Section \ref{Section_Evolution of the tensor modes}, we analyze the evolution of the tensor modes of the fluid and the gravitational waves, where we present the main results of this paper. Finally, Section \ref{Section_Conclusions} contains our conclusions.

Appendix \ref{Section_Derivation of Hydro equations} provides the derivation of the hydrodynamical equations from kinetic theory. Appendix \ref{Section_L functions} describes the functions that appear in the analysis of the hydrodynamical model. In Appendix \ref{Section_Relaxation time calculation}, we evaluate the relaxation time in our model.

\section{The model}
\label{Section_The model}
We adopt a field-theoretic description of the inflaton field in Minkowski spacetime with metric signature $(-,+,+,+)$, governed by the usual Klein–Gordon equation, and assume that the field forms a classical homogeneous condensate. We model its fluctuations, and all other relevant forms of matter present, as a relativistic causal fluid, and describe its dynamics using a second order theory within the so-called ``divergence type theory''  class (DTT) \cite{geroch_1990,1986_Liu,2020_Cantarutti_Calzetta}. The detailed derivation of the hydrodynamic equations is presented in Appendix~\ref{Section_Derivation of Hydro equations}. The fluid and the condensate couple through the fluid effective mass \cite{2008_Calzetta_Hu,2021_Hindmarsh}, which is defined by a gap equation, see Eq. (\ref{Gap equation fluid mass}).

The basic issue in the theory of relativistic real fluids is that energy-momentum conservation provides only four equations for the ten components of the energy-momentum tensor (since we assume zero chemical potential, because of matter-antimatter symmetry, there is no independent law of particle number conservation). In the DTT framework this gap is filled by introducing further equations, which take the form of conservation laws for a number of 'nonequilibrium tensors'. In this paper we shall introduce two such tensors, a third rank one $A^{\mu\nu\rho}$ and a fourth rank one $A^{\mu\nu\rho\sigma}$. This is a minimal setup which enables the propagation of tensor waves in the fluid \cite{2021_Perna_Calzetta}. The complete set of hydrodynamic equations take the form
\bea
\label{Conservation laws}
\nabla_\nu T^{\mu \nu}_f &= F^\mu M_T^2\\\
S_{\mu\nu}^{\alpha\beta}\left[A^{\mu \nu \rho}_{ ; \rho} - A^{\mu \nu \rho \sigma} u_{\rho ; \sigma} - I^{\mu \nu}\right] &= S_{\mu\nu}^{\alpha\beta}\left[2F^\mu B^\nu + F^\sigma u_\sigma B^{\mu \nu}\right]\\
S_{\mu\nu\rho}^{\alpha\beta\gamma}\left[A^{\mu \nu \rho \sigma}_{ ; \sigma} - 2A^{\mu \nu \rho \sigma\lambda} u_{\sigma ; \lambda} - I^{\mu \nu \rho}\right] &= S_{\mu\nu\rho}^{\alpha\beta\gamma}\left[3F^\mu B^{\nu \rho} + 2F^\sigma u_\sigma B^{\mu \nu \rho}\right].
\tea
Symmetry and dimensional arguments alone are not capable of singling out a whole set of constituve relations for the tensors in these equations. To give content to these equations, therefore, it is necessary to derive them from an underlying theory. For example, we may assume that underlying the hydrodynamic description there is a description of the system as quasiparticle excitations with a well defined mass. Then we may postulate a kinetic theory description for those quasiparticles and derive hydrodynamics from it. This is carried out in Appendix \ref{Section_Derivation of Hydro equations}. The resulting constitutive relations are given in eqs. (\ref{Tensors in the conservation laws}), (\ref{collision integral}) and (\ref{Boltzmann external force}).

The fluid energy-momentum tensor $T_f^{\mu\nu}$ defines the four-velocity $u^\mu$ and energy density $\epsilon$ via the Landau prescription $T_f^{\mu\nu} u_\nu = -\epsilon u^\mu$. The energy density $\epsilon$ then determines the fluid temperature $T$ through the equilibrium equation of state. Together, $u^\mu$ and $T$ define the inverse temperature vector $\beta^\mu = u^\mu / T$.

In Eq. \eqref{Conservation laws} we defined the tensors $S_{\mu\nu}^{\alpha\beta}$ and $S_{\mu\nu\rho}^{\alpha\beta\gamma}$, which are projectors onto the transverse and traceless components

\bea
\label{Transverse and traceless projectors}
S_{\mu\nu}^{\alpha\beta}&=\frac12\left\{
\Delta_{\mu}^\alpha \Delta_{\nu}^\beta+\Delta_{\mu}^\beta \Delta_{\nu}^\alpha
-\frac{2}{3}\Delta^{\alpha\beta}\Delta_{\mu\nu}\right\}\\
S_{\mu\nu\rho}^{\alpha\beta\gamma}&=\frac16\Big\{
\Delta_{\mu}^{\alpha} \Delta_{\nu}^{\beta} \Delta_{\rho}^{\gamma}
+ \Delta_{\mu}^{\alpha} \Delta_{\rho}^{\beta} \Delta_{\nu}^{\gamma}
+ \Delta_{\nu}^{\alpha} \Delta_{\mu}^{\beta} \Delta_{\rho}^{\gamma} \\
&\,\,\quad + \Delta_{\nu}^{\alpha} \Delta_{\rho}^{\beta} \Delta_{\mu}^{\gamma}
+ \Delta_{\rho}^{\alpha} \Delta_{\mu}^{\beta} \Delta_{\nu}^{\gamma}
+ \Delta_{\rho}^{\alpha} \Delta_{\nu}^{\beta} \Delta_{\mu}^{\gamma}\\
&\,\,\quad-\frac{2}{5}\Big[
\Delta^{\alpha\beta}\lc
\Delta^\gamma_\mu \Delta_{\nu\rho} +
\Delta^\gamma_\nu \Delta_{\mu\rho} +
\Delta^\gamma_\rho \Delta_{\mu\nu}\rc\\
&\,\,\,\,\,\,\,\,\;\quad+
\Delta^{\beta\gamma}\lc\Delta^\alpha_\mu \Delta_{\nu\rho} +
\Delta^\alpha_\nu \Delta_{\mu\rho} +
\Delta^\alpha_\rho \Delta_{\mu\nu}\rc\\
&\,\,\,\,\,\,\,\,\;\quad+
\Delta^{\gamma\alpha}\lc\Delta^\beta_\mu \Delta_{\nu\rho} +
\Delta^\beta_\nu \Delta_{\mu\rho} +
\Delta^\beta_\rho \Delta_{\mu\nu}\rc\Big]\Big\}
\tea
where $\Delta^{\mu\nu}=u^\mu u^\nu+g^{\mu\nu}$ as usual. They make sure the system is not overdetermined.

The tensors $I^{\mu \nu}$ and $I^{\mu \nu\rho}$ encode the dissipative effects. Then, the entropy production takes the form
\be 
\label{entropy_production}
S^{\mu}_{;\mu}=\zeta_{\mu\nu}I^{\mu \nu}+\xi_{\mu\nu\rho}I^{\mu \nu\rho}\ge 0.
\te
The new tensors $\zeta_{\mu\nu}$ and $\xi_{\mu\nu\rho}$, together with the inverse temperature vector, are the degrees of freedom of the theory. These nonequilibrium tensors are fully symmetric, transverse and traceless
\bea
\label{Properties nonequilibrium tensors}
\zeta_{\mu\nu}u^\mu=\zeta^\mu_\mu &=0\\
\xi_{\mu\nu\rho}u^\mu=\xi^{\mu\nu}_\mu &=0
\tea
and therefore invariant under the transverse and traceless projectors. 

By definition, these nonequilibrium tensors vanish identically in equilibrium. We define the linearized theory around local equilibrium by assuming constitutive laws that respect the symmetries of the various tensors and are linear in $\zeta_{\mu\nu}$ and $\xi_{\mu\nu\rho}$. Substituting these constitutive laws into the conservation equations, we obtain
\bea
\label{Dynamical equations fluid}
T_{0;\nu}^{\mu\nu}+\nabla_\nu\lc T_1^{\mu\nu\alpha\beta}\zeta_{\alpha\beta} \rc &=F^\mu M_T^2
\\
S^{\alpha\beta}_{\mu\nu}\lc T^{\mu\nu\rho}_{1\,;\rho}+\left[T_2^{\mu\nu\rho\sigma\lambda}\zeta_{\sigma\lambda}+T_3^{\mu\nu\rho\sigma\lambda\tau}\xi_{\sigma\lambda\tau}\right]_{;\rho}-T_2^{\mu\nu\rho\sigma}u_{\rho;\sigma}+\frac{1}{\tau}T_1^{\mu\nu\rho\sigma}\zeta_{\rho\sigma} \rc
&=S^{\alpha\beta}_{\mu\nu}\lc F^\lambda u_\lambda T_3^{\mu\nu\rho\sigma}\zeta_{\rho\sigma} \rc
\\
S_{\mu\nu\rho}^{\alpha\beta\gamma}\lc 
\left[T_3^{\mu\nu\rho\sigma\lambda\tau}\zeta_{\lambda\tau}
+T_4^{\mu\nu\rho\sigma\lambda\tau\delta}\xi_{\lambda\tau\delta}\right]_{;\sigma}
+\frac1\tau T_3^{\mu\nu\rho\sigma\lambda\tau}\xi_{\sigma\lambda\tau} \rc
&=S_{\mu\nu\rho}^{\alpha\beta\gamma}\big( 3F^\mu T_3^{\nu\rho\sigma\lambda}\zeta_{\sigma\lambda}
\\
&+2F^\lambda u_\lambda
T_5^{\mu\nu\rho\sigma\tau\delta}\xi_{\sigma\tau\delta} \big)
\tea
where we defined
\bea
\label{General tensors structure}
T_m^{\mu_1 \cdots \mu_n}=&\,C^n_m(0)\; u^{\mu_1}\cdots u^{\mu_n}\\
+&\,C^n_m(1)\left[ \Delta^{\mu_1\mu_2}u^{\mu_3}\cdots u^{\mu_n} +\cdots\right]\\
+&\,\cdots\\
+&\,C^{n}_m(\lfloor n/2 \rfloor)[\cdots].
\tea
Here $\lfloor x \rfloor$ is the floor function introduced because if the tensor has an even number of indices, the last term contains $n/2$ projectors $\Delta^{\mu\nu}$, whereas if it has an odd number of indices, it contains $(n-1)/2$ projectors and one four-velocity.

Hydrodynamics alone cannot determine the actual values of the $C^n_m(k)$ coefficients. In Appendix~\ref{Section_Derivation of Hydro equations} we show that the DTT model can be derived from kinetic theory, from which we obtain
\be
\label{C_functions}
C^n_m(k) = \frac{1}{(2k+1)!!} a^{2k}_{n-m-2k},
\te 
where we define the functions
\be
\label{a^k_l functions}
a^k_l = \int Dp \, |\mathbf{p}|^k (p^0)^l f_0,
\te
with the decomposition of the four-momentum as $p^\mu = (p^0, \mathbf{p})$. We linearize the theory around a state in which the fluid is at rest, $u^\mu = (1,0,0,0)$, with a time-dependent temperature, consistent with coupling to a homogeneous condensate. The function $f_0$ is the local equilibrium distribution
\be
\label{local_equilibrium_distribution}
f_0 = e^{-\frac{p^0}{T}} = e^{-\frac{1}{T}\sqrt{\mathbf{p}^2 + M^2}},
\te
corresponding to the relativistic Maxwell-Boltzmann case, with $M$ the fluid mass. The invariant measure $Dp$ is
\be
\label{Invariant measure}
Dp = 2 n_* \frac{d^4p}{(2\pi)^3} \delta(p^2 + M^2) \theta(p^0) = n_* \frac{d^3p}{(2\pi)^3} \frac{1}{p^0},
\te
where $n_*$ is the number of degrees of freedom of the fluid.

The first line in \eqref{Conservation laws} describes the non-conservation of the fluid energy-momentum tensor, which, when expanded up to first order in the nonequilibrium tensors, reads
\be
\label{Perturbed fluid EMT}
T_f^{\mu\nu}=a^0_2 u^\mu u^\nu+\frac{1}{3}a^2_0 \Delta^{\mu\nu}+\frac{2}{15}a^4_{-1}\zeta^{\mu\nu}
\te
where the first two terms correspond to the equilibrium energy–momentum tensor, $T_0^{\mu\nu}$, and the third term represents the viscous contribution, $\Pi^{\mu\nu}$, that is induced due to the nonequilibrium tensor $\zeta^{\mu\nu}$. This non-conservation is a consequence of the external force acting on the fluid due to the interaction with the condensate. Nevertheless, the total energy-momentum tensor must be conserved $\nabla_\nu(T_f^{\mu\nu}+T_\phi^{\mu\nu})=0$, which leads to the non-conservation of the condensate energy-momentum tensor
\be
\label{Inflaton EMT non-conservation}
\nabla_\nu T_\phi^{\mu\nu}=-F^\mu M_T^2.
\te
Since we are working with linear perturbations out of equilibrium, the thermal mass $M_T^2$ is evaluated using the equilibrium distribution function
\be 
\label{thermal_mass}
M_T^2 = \int Dp \, f_0.
\te
The external force that couples the fluid to the condensate is given by \cite{2008_Calzetta_Hu,2021_Hindmarsh}
\be
\label{Boltzmann external force}
F_\mu=-\frac12 \partial_\mu M^2.
\te

In this framework, the Klein--Gordon equation must be modified by the addition of an external source term, yielding  
\be
\label{Modified Klein-Gordon equation}
-\Box\phi + V'(\phi) = -K \,,
\te
where $K$ depends on both the condensate and the fluid. The energy--momentum tensor of the condensate is
\be
\label{Inflaton EMT}
T^\phi_{\mu\nu} = \phi_{,\mu}\phi_{,\nu} - g_{\mu\nu}\left[ \tfrac12 g^{\alpha\beta}\phi_{,\alpha}\phi_{,\beta} + V(\phi)\right]  \,
\te
We omit a possible cosmological constant term, which is consistent with the symmetries but not required. Substituting this form into Eq.~(\ref{Inflaton EMT non-conservation}) leads to the following consistency relation
\be
\label{Consistency equation}
\tfrac12 M^2_{,\mu} M_T^2 = K \phi_{,\mu} \,.
\te
To close the system of equations, we must introduce a gap equation for the fluid mass \cite{2008_Calzetta_Hu}, which we take to be  
\be
\label{Gap equation fluid mass}
M^2 = m^2 + g_{cf}^2 \phi^2 \,,
\te
where $m$ is the bare mass of the fluid and $g_{cf}$ denotes the condensate--fluid coupling constant. This effective mass has the same form as in the standard theory of reheating with two scalar fields \cite{2019_Lozanov}. Therefore, Eq. (\ref{Consistency equation}) reduces to

\be
\label{K function integrated}
K(\phi,z)=g_{cf}^2 M_T^2\phi.
\te

\section{Background dynamics}
\label{Section_background}
We begin by studying the background dynamics, whose degrees of freedom are the homogeneous condensate and fluid temperature. We consider a quartic inflaton potential $V(\phi)=\frac14\lambda\phi^4$. The background equations consist of the modified Klein-Gordon equation for the condensate \eqref{Modified Klein-Gordon equation}, and the non-conservation of the energy-momentum tensor for the fluid 

\be
\label{First Boltzmann equation}
\dot\epsilon=\frac12 M_T^2\partial_t M^2
\te
where $\epsilon=a^0_2$ denotes the Landau-Lifshitz energy density of the fluid. This equation governs the evolution of the fluid temperature. 

From the structure of equation \eqref{First Boltzmann equation} and the inspection of the coefficients \eqref{a^k_l functions}, we observe that using $\phi$ and $z = M/T$ as independent variables, rather than $\phi$ and $T$, simplifies the analysis. This allows us to factor out the dependence on each independent variable in the coefficients $a^k_l$
\be
\label{M z decomposition of coefficients}
a^k_l(M,z)=\frac{n_*}{2\pi^2}M^{k+l+2}L_{k,l}(z)
\te
where we define the functions
\be
L_{k,l}(z)=\int_0^\infty dx\,\sinh^{k+2}(x)\,\cosh^l(x) \, e^{-z\cosh x}.
\te
Appendix \ref{Section_L functions} details the structure of these functions. 

\begin{figure}
    \centering
    \includegraphics[width=.65\linewidth]{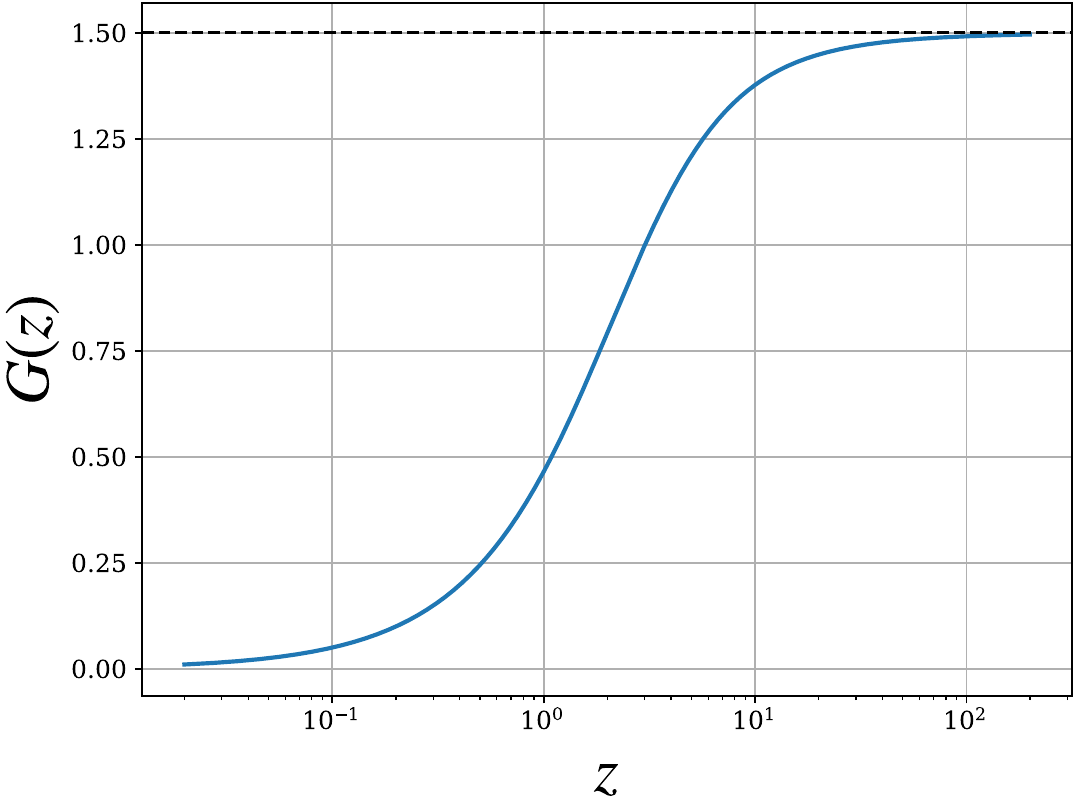}
    \caption{Graph of the function $G(z)$, defined in \eqref{G(z) function}, that appears in the background equation of the fluid \eqref{Fluid background equation simplified}. It exhibits two simple asymptotic regimes: for $z\ll1$ we have $G(z)\approx z/2$ and, for $z\gg1$, $G(z)\approx 3/2$. In between these two regimes there is a fast transition.}
    \label{fig:G(z) graph}
\end{figure}

Eq. (\ref{First Boltzmann equation}) may be rewritten using equation \eqref{M z decomposition of coefficients} for the coefficients $a^k_l$ yielding
\be
\label{Fluid background equation simplified}
\dot z=\frac{M^2_{,t}}{M^2}G(z)
\te
where
\be
\label{G(z) function}
G(z)=\frac{2L_{0,2}(z)-\frac12 L_{0,0}(z)}{L_{0,3}(z)}=\frac32 \frac{K_3(z)}{3K_3(z)/z+K_2(z)}.
\te
This function $G(z)$, despite being a rather involved quotient of Bessel functions $K_\nu(z)$, exhibits simple behavior for most values of $z$, as can be seen in Fig. \ref{fig:G(z) graph}. For $z\gtrsim10$, this function approaches a constant $G\approx\frac32$ and for $z\lesssim1$, it becomes $G\approx \frac{z}{2}$. In the intermediate range the function has a transition between both regimes. In this work we are interested in the first of these regimes, $z\gtrsim10$, where we can use the asymptotic expansion of the Bessel functions to obtain
\be
\label{z(t) for big z}
z(t)=z_0+3 \ln\lc\frac{M(t)}{M_0}\rc
\te
where $z_0$ is the initial condition for $z$, and $M_0$ the initial fluid mass. In this case the fluid temperature $T=M/z$ oscillates together with the fluid mass. This solution gives an explicit relation $z(M)$ which may be expressed in terms of the condensate $\phi(t)$ through the gap equation for the fluid mass \eqref{Gap equation fluid mass}.

Inflation in the theory with potential $V=\frac{\lambda}{4}\phi^4$ ends when the condensate reaches $\phi_0\sim m_p$, so considering this value as the initial condition for the inflaton field at the beginning of the reheating period, the mass ratio in \eqref{z(t) for big z} satisfies $(1+g_{cf}^2\phi_0^2/m^2)^{-1}<M/M_0<1$. We see that if $g_{cf}^2\phi_0^2/m^2\ll1$ and $z_0\gtrsim 10$, the time variation of $z$ becomes negligible.

On the other hand, the modified Klein-Gordon equation becomes, using the expression \eqref{K function integrated},
\be
\label{Final condensate equation}
\ddot{\phi}+\lc m_r^2(z)+\lambda_r(z)\,\phi^2\rc \phi=0
\te
with
\bea
\label{Renormalized inflaton couplings}
m_r^2(z)&=\frac{n_*}{2\pi^2} m^2 g_{cf}^2 L_{0,0}(z)\\
\lambda_r (z)&=\lambda+\frac{n_*}{2\pi^2} g_{cf}^4 L_{0,0}(z).
\tea
We see that the condensate-fluid coupling induces an effective mass for the condensate and renormalizes the self-coupling constant $\lambda$. These corrections decay exponentially for $z\gg1$, thus in this regime we expect the condensate to undergo the usual dynamics of Reheating with potential $V\sim\lambda\phi^4$ \cite{1997_Greene,2018_Lozanov}. Note that if we had started with a quadratic potential, the fluid would have produced the same effect: a renormalization of the bare mass of the condensate and the introduction of a qubic term in the equation of motion emulating a renormalized $\lambda \phi^4$ potential.

At the end of the inflationary era, the beginning of Reheating, all the energy of the universe was contained in the inflaton $\phi$, so the fluid temperature must be much smaller than the condensate amplitude, $T_0\ll\phi_0$, which implies $z_0\gg1$. In this work, we assume this is the case and we take $z\gg1$ as constant. This approximation should hold as long as the backreaction from the fluid to the condensate remains negligible. Under this assumption, $m_\phi$ and $\lambda_r$ in \eqref{Renormalized inflaton couplings} remain constant, and we are left with the KG equation
\be
\ddot{\phi}+\lc m_r^2+\lambda_r\,\phi^2\rc \phi=0
\te
where the renormalized couplings may depend on $z$. Since we are working in the linear regime of the reheating period, we use the Hartree approximation to get a periodic solution to this equation, which is valid for a few dozen oscillations—the typical duration of the linear stage of Reheating \cite{1997_KLS,2019_Lozanov}. The resulting solution is $\phi(t)=\phi_0\cos(\Omega t)$, with
\be
\label{Omega Hartree}
\Omega^2=m_r^2+\frac34 \lambda_r \phi_0^2
\te
where $\phi_0$ is the initial amplitude of the condensate, corresponding to the value of the field at the end of inflation.

In summary, the background dynamics in the case of interest can be described as
\bea
\begin{cases}
   z(t)\simeq z_0\sim {\rm const.}\\
   \phi(t)\simeq \phi_0 \cos\lc \Omega t\rc\,,
\end{cases}
\tea
where $\Omega^2=m_r^2(z_0)+3\lambda_r(z_0)\phi_0^2/4\sim$ const. From now on, we write the initial value $z_0$ simply as $z$.

\section{Tensor perturbation dynamics}
\label{Section_Tensor perturbation dynamics}

In this work we are only interested in the tensor perturbations of the fluid, which will source the gravitational waves. We consider a fluctuation in the metric $g_{\mu\nu}=\eta_{\mu\nu}+h_{\mu\nu}$ where, as usual, $h_{\mu\nu}$ is a transverse and traceless spatial tensor that describes gravitational waves. The remaining equations correspond to the Einstein equation for gravitational waves \cite{2018_Maggiore,Schutz_2022,2018_Caprini} and to the equations (\ref{Dynamical equations fluid}) for the nonequilibrium tensor perturbations. The full system of equations is then (see Appendix \ref{Section_Derivation of Hydro equations})
\bea
\label{Second and third Boltzmann equation and Einstein equation}
\dot\zeta^{ij}+\left(\frac1{\tau}+\frac{\Dot{a}^{4}_{-1}}{a^{4}_{-1}}+\frac12\frac{a^4_{-3}}{a^4_{-1}}M^2_{,t}\right)\zeta^{ij}+\frac37\frac{a^6_{-3}}{a^4_{-1}}\xi^{ijk}_{,k}
&=-\frac{5a^2_{0}-a^4_{-2}}{2a^4_{-1}}\dot h^{ij}\\
\dot\xi^{ijk}+\left(\frac1{\tau}+\frac{\Dot{a}^{6}_{-3}}{a^{6}_{-3}}+\frac{a^6_{-5}}{a^6_{-3}}M^2_{,t}\right)\xi^{ijk}+\frac13\left(\zeta^{ij,k}+\zeta^{jk,i}+\zeta^{ki,j}\right)&=0\\
\ddot h^{ij}-\nabla^2 h^{ij}&=\frac{4}{15}\frac{a^{4}_{-1}}{m_p^2}\zeta^{ij}.
\tea
We assume that the variable $z$ is constant, and that the condensate exhibits periodic oscillations of the form $\phi = \phi_0 \cos(\Omega t)$, as discussed in the previous section. We use the standard decomposition for the tensor degrees of freedom in Fourier space
\bea
\zeta_{ij}(\vec x,t)&= \int \frac{ d^3k}{(2\pi)^3}\,e^{i\mathbf{k}\cdot\mathbf{x}}\,\sum_\sigma \zeta_{\mathbf{k},\sigma}(t)\,\epsilon^\sigma_{ij}(\hat k)\\
h_{ij}(\vec x,t)&= \int \frac{ d^3k}{(2\pi)^3}\,e^{i\mathbf{k}\cdot\mathbf{x}}\,\sum_\sigma h_{\mathbf{k},\sigma}(t)\,\epsilon^\sigma_{ij}(\hat k)\\
\xi_{ijk}(\vec x,t)&=\int \frac{ d^3k}{(2\pi)^3}\,e^{i\mathbf{k}\cdot\mathbf{x}}\sum_\sigma i \, \xi_{\mathbf{k},\sigma}(t)
\Big[k_i\,\epsilon^\sigma_{jk}(\hat k)+k_j\,\epsilon^\sigma_{ki}(\hat k)+k_k\,\epsilon^\sigma_{ij}(\hat k)\Big]\label{tensor decomposition}
\tea
where $\sigma$ corresponds to the polarization type. Here $\epsilon^\sigma_{ij}(\hat k)$ is the polarization tensor, which is transverse and traceless: $\epsilon^\sigma_{ij}(\hat k)k_i = 0$ and $\epsilon^\sigma_{ii}(\hat k) = 0$. It also satisfies the normalization condition $\epsilon^\sigma_{ij}(\hat k)\,\epsilon^{\sigma'*}_{ij}(\hat k)=2\delta_{\sigma\sigma'}$. From this point onward, we drop the polarization index, since no mechanism is assumed to distinguish between them.

The coefficients in the dynamical equations for the perturbations \eqref{Second and third Boltzmann equation and Einstein equation} contain terms with factors of $M_{,t}$ and $M$. The first one crosses from positive to negative values on each oscillation of the condensate, while the latter is bounded away from zero. To simplify the analysis we approximate the oscillatory fluid mass $M$ in Eq.~\eqref{Gap equation fluid mass} by its root-mean-square (rms) value $M \approx \sqrt{\langle M^2 \rangle} = m \, \sqrt{1 +  q/2}$, where $q = g_{cf}^2 \phi_0^2/m^2$. This provides a reliable approximation as long as $q \lesssim 2$ and $z\gtrsim1$. Therefore, we obtain
\be
\frac{M^2_{,t}}{M^2} \simeq  -\Omega \, \frac{q \, \sin(2\Omega t)}{1 + q/2} \,.
\te

The equations of motion also involve the relaxation time from the collision integral \eqref{collision integral}, whose explicit expression, derived in Appendix~\ref{Section_Relaxation time calculation}, is
\be
\label{Relaxation time}
\tau = \lc\frac{\eta}{s}\rc \frac{3}{M} \, \frac{K_3(z)}{L_{2,0}(z) - \tfrac15 L_{4,-2}(z)} \,,
\te
with $\eta/s$ the ratio between the shear viscosity and the entropy density of the fluid. In this case, this is a free parameter depending on the type of fluid interaction that takes values from $\eta/s>1/4\pi$, where the lower bound comes from AdS/CFT analysis in the strong coupling limit \cite{2001_Policastro,2001_Cremonini}. On the other hand, a representative value for a weakly coupled quark-gluon plasma in QCD is $\eta/s\sim10^3$ \cite{2020_McDonough}. Here again we obtain a constant relaxation time after replacing the fluid mass by its rms value. The asymptotic behavior of the inverse relaxation time, in the regime $z\gg1$, is
\be
\label{Tau for big z}
\frac1\tau \simeq \lc\frac{\eta}{s}\rc^{-1}\frac{M}{z^2}=\lc\frac{\eta}{s}\rc^{-1}\frac{m\sqrt{1+q/2}}{z^2}.
\te

Based on the previous discussion, the equations of motion \eqref{Second and third Boltzmann equation and Einstein equation} for the tensor modes $\zeta_k$, $\xi_k$ and $h_k$ in \eqref{tensor decomposition} read
\begin{align}
\dot\zeta_\mathbf{k} + \Big(\frac{1}{\tau} - 2\Omega\sqrt{b}\,\sin{2\Omega t}\Big)\zeta_\mathbf{k}
    - \frac{3}{7}\frac{k^2}{r}\,\xi_\mathbf{k} &= -\frac{s}{2}\,\dot h_\mathbf{k} \,, 
    \label{Perturbation equations simplified 1} \\
\dot\xi_\mathbf{k} + \Big(\frac{1}{\tau} - 2 \Omega p \sqrt{b}\,\sin{2\Omega t}\Big)\xi_\mathbf{k}
    + \frac{1}{3}\zeta_\mathbf{k} &= 0 \,,
    \label{Perturbation equations simplified 2} \\
\ddot h_\mathbf{k} + k^2 h_\mathbf{k} &= u \zeta_\mathbf{k} \,.
    \label{Perturbation equations simplified 3}
\end{align}
where the (constant) coefficients are defined as
\bea
\label{eq_coefficients}
\sqrt b&=\frac{q}{4(1+q/2)}\lc 5+\frac{L_{4,-3}(z)}{L_{4,-1}(z)}\rc\\
p&=\lc 5+2\frac{L_{6,-5}(z)}{L_{6,-3}(z)}\rc\lc 5+\frac{L_{4,-3}(z)}{L_{4,-1}(z)}\rc^{-1}\\
r&=\frac{L_{4,-1}(z)}{L_{6,-3}(z)}\\
s&=\frac{1}{M}\frac{5\,L_{2,0}(z)-L_{4,-2}(z)}{L_{4,-1}(z)}\\
u&=\frac{2n_* M^5}{15\pi^2 m_p^2}L_{4,-1}(z)\,.
\tea
The properties of the functions $L_{k,l}$ imply that $b\leq9$, $p\in(1,7/6)$ and $r\geq1$.

We analyze the system \eqref{Perturbation equations simplified 1}--\eqref{Perturbation equations simplified 3} neglecting both the backreaction of the fluid on the condensate and the backreaction of the gravitational waves on the fluid. This corresponds to the regime $su/\Omega^4 \ll 1$ which holds for the case of interest $z \gtrsim 10$, thus we set $s=0$ in \eqref{Perturbation equations simplified 1}. In this way, the fluid evolves independently and sources the production of gravitational waves through Eq. \eqref{Perturbation equations simplified 3}.

In particular, we are interested in the production of gravitational waves sourced by the tensor fluctuations of the fluid. For this reason, we impose non-trivial initial conditions on the fluid tensor modes ($\zeta_k$ and $\xi_k$), while the gravitational waves ($h_k$) are taken to have vanishing initial conditions.

Since both the tensor perturbations of the fluid and the gravitational waves are treated as stochastic variables \cite{2018_Caprini}, we follow the standard approach~\cite{2021_Miron-Granese,2007_Dufaux} in which the relevant observables are their spectra, or equivalently, their correlation functions. To make this explicit, each dynamical variable is decomposed into a stochastic initial condition and a deterministic evolution factor. In this work we shall only consider initial fluctuations for the second-rank nonequilibrium tensor $\zeta_{\mu\nu}$, which means that the Fourier mode $\xi_\mathbf{k}(t)$ has trivial initial conditions $\xi_\mathbf{k}(0)=0$. This allows us to do the following decomposition
\bea
\label{Stochastic+deterministic Fourier modes}
\zeta_\mathbf{k}(t) &= \zeta_{\mathbf{k},\sigma}^{\rm ini}\,\tilde\zeta_k(t) , \\
\xi_\mathbf{k}(t)   &= \zeta_{\mathbf{k},\sigma}^{\rm ini}\,\Omega^{-1}\tilde\xi_k(t)   , \\
h_\mathbf{k}(t)     &= \zeta_{\mathbf{k},\sigma}^{\rm ini}\,\Omega\,g_k(t).
\tea
where $\tilde\zeta_k$, $\tilde\xi_k$ and $g_k$ contain the deterministic evolution and are all dimensionless. Then, the stochastic initial conditions are solely characterized by the spectrum of $\zeta_{\mathbf{k},\sigma}^{\rm ini}$, which is defined as
\be
\label{Initial spectra}
\langle \zeta_{\mathbf{k},\sigma}^{\rm ini}\,\zeta_{\mathbf{k}',\sigma'}^{\rm ini}\rangle = (2\pi)^3\delta_{\sigma\sigma'}\delta(\mathbf{k}+\mathbf{k}')\mathcal{P}_\zeta(k).
\te
We shall further discuss the spectrum $\mathcal{P}_\zeta(k)$ in Section \ref{Section Initial conditions}.

Since only the second-rank tensor $\zeta_{\mu\nu}$ contributes to the production of gravitational waves, we shall reduce the two first-order equations for $\zeta_{\mu\nu}$ and $\xi_{\mu\nu\rho}$ to a single second-order equation for the evolution function $\tilde\zeta_k(t)$. The resulting equation contains a dissipative term, which may be elliminated through the substitution
\bea
\label{Dissipative term substitution}
\tilde\zeta_\kappa(\theta) &= y_\kappa(\theta)\,\exp\!\left[-\int^\theta d\theta' 
\left( \frac{1}{\tilde{\tau}} - (1+p)\sqrt{b}\,\sin 2\theta' \right)\right] \\
&= y_\kappa(\theta)\,\exp\!\left[-\frac{\theta}{\tilde{\tau}} 
+ (1+p)\sqrt{b}\,\sin^2\theta \right].
\tea
Here, to simplify the notation, we introduced the dimensionless variables  
$\theta = \Omega t$, $\kappa = k/\Omega$, and $\tilde{\tau} = \Omega \tau$, 
with primes denoting derivatives with respect to $\theta$. Note that in $\tilde\zeta_\kappa$, in contrast to $y_\kappa$, there is an exponentially decaying factor, which corresponds to the dissipation due to the collision integral, together with an oscillatory factor.

Consequently, under all these considerations, equations \eqref{Perturbation equations simplified 1} and \eqref{Perturbation equations simplified 2} give a second-order wave equation for $y_\kappa$, namely a Mathieu-like equation \cite{2012_Teschl}
\be
\label{Mathieu-like equation}
y''_\kappa+\omega^2_\kappa \,\, y_\kappa=0\,,
\te
where the time-dependent frequency is
\be
\label{Mathieu-like equation frequency}
\omega^2_\kappa(\theta)=\frac{\kappa^2}{7r}-\frac12\delta^2+2\delta\cos{2\theta}+\frac12\delta^2 \cos{4\theta}\,
\te
and $\delta=\sqrt{b}(p-1)$ is the resonance parameter. Note that this resonance parameter satisfies $0 \leq \delta \leq \frac{1}{2}$. From the theory of the Mathieu equation, we expect the variable $y_\kappa$ to undergo a process of parametric resonance in which a band of momenta is exponentially amplified, roughly as $y_\kappa\sim e^{\tilde\mu_\kappa \theta}$ with $\tilde\mu_\kappa$ the Floquet exponent. Moreover, since the resonance parameter is small, the resonance will be narrow. In this sense, the amplification of the variable $y_\kappa$ will be similar to that occurring for scalar fields coupled to the inflaton in the standard theory of reheating \cite{2019_Lozanov}.

We also use the substitution \eqref{Dissipative term substitution} in the Einstein equation for the gravitational waves \eqref{Perturbation equations simplified 3}, to finally get the dynamical system for the evolution functions $y_\kappa$ and $g_\kappa$, namely
\begin{empheq}[left=\empheqlbrace]{align}
y''_\kappa + \omega^2_\kappa \, y_\kappa &= 0 \,, 
   \label{eq:y-eq} \\
g''_\kappa + \kappa^2 g_\kappa &= \frac{u}{\Omega^3} \, y_\kappa \,
   e^{-\tfrac{\theta}{\tilde{\tau}} + (1+p)\sqrt{b}\,\sin^2\theta} \,,
   \label{Einstein equation for GW (final)}
\end{empheq}
where $\omega_\kappa$ is the time-dependent frequency given by \eqref{Mathieu-like equation frequency} and $u/\Omega^3$ is a dimensionless parameter acting as an effective coupling between the fluid and the gravitational waves.

The last step is to specify the initial conditions for the Fourier modes introduced in~\eqref{Stochastic+deterministic Fourier modes}.  
As stated earlier, we assume vanishing initial conditions for the third-rank nonequilibrium tensor $\xi_{\mu\nu\rho}$ and the gravitational waves. Consequently, from~\eqref{Perturbation equations simplified 1}, the Fourier modes of the second-rank tensor and the gravitational waves satisfy $\dot{\zeta}_k(0) = -\frac{1}{\tau}\,\zeta_k(0)$ and $h_k(0) = 0,\; \dot{h}_k(0) = 0$ respectively.

In this way, the deterministic evolution functions in \eqref{eq:y-eq}-\eqref{Einstein equation for GW (final)} initialize according to
\begin{align}
y_\kappa(0) &= 1\,, \quad y'_\kappa(0) = 0\,, \\
g_\kappa(0) &= 0\,, \quad g'_\kappa(0) = 0\,.
\end{align}

Finally, to fully specify the initial conditions of the physically relevant fluctuations in \eqref{Stochastic+deterministic Fourier modes}, we require the spectrum of stochastic initial values, $\langle \zeta_{k,\sigma}^{\rm ini}\zeta_{k',\sigma'}^{\rm ini}\rangle$, whose derivation is presented in Section~\ref{Section Initial conditions}.

\section{Spectra of perturbations}
\label{Section_Spectras}

In this section we analyze the spectra relevant for the study of the amplification of the fluid and the gravitational waves. In the first subsection, we derive the initial conditions of the stochastic components of the nonequilibrium tensors defined in \eqref{Stochastic+deterministic Fourier modes}. In the second subsection, we compute the energy density spectra of the fluid and the gravitational waves and assess the validity of the linear approach adopted in this work. Finally, we provide a rough order-of-magnitude estimate of the amplification of the energy fraction in gravitational waves, which is the most relevant observable in this study.

\subsection{Stochastic initial fluctuations}
\label{Section Initial conditions}

To determine the initial conditions for the tensor perturbations of the fluid, we shall assume this is in equilibrium at the beginning of reheating, thus subject to hydrodynamic fluctuations \cite{Lifshitz1980StatPhys2,Calzetta_1998,2022_Miron-Granese,Miron-Granese_jhep_2020}. These can be obtained by adding a noise term to the Boltzmann equation, which was used to derive the hydrodynamical model in Appendix~\ref{Section_Derivation of Hydro equations}, with a statistics determined by the fluctuation-dissipation theorem. In equilibrium, the hydrodynamic fluctuations have a probability density function of the large deviation theory type $e^{\Phi}$ \cite{Touchette_2009}, where
\be
\Phi=\int d^3x\, u_\mu \Phi^\mu
\te
and $\Phi^{\mu}$ is the generating function of the divergence type theory \cite{2020_Cantarutti_Calzetta}, namely
\be
\Phi^\mu=S^\mu+\beta_\nu T^{\mu\nu}+\zeta_{\nu\rho}A^{\mu\nu\rho}+\xi_{\nu\rho\sigma}A^{\mu\nu\rho\sigma}
\te
with $S^\mu$ the entropy current \eqref{entropyflux}. Observe that the fluctuations of the second and third rank tensors $\zeta_{\mu\nu}$ and $\xi_{\mu\nu\rho}$ are stochastically independent. For simplicity  we only consider initial fluctuations for the second-rank tensor, whereby, to quadratic order, we obtain 
\be
\Delta\Phi=-\frac12\int d^3x\, 
\frac{2}{15}a^4_{-1}\zeta^{\mu\nu}\zeta_{\mu\nu}
\,.
\te

The initial probability density is gaussian and the correlation function is
\be
\langle \zeta^{\mu\nu}(\mathbf{x},0)\zeta_{\alpha\beta}(\mathbf{x}',0) \rangle =
\frac{15}{2 a^4_{-1}} S^{\mu\nu}_{\alpha\beta}\,\delta(\mathbf{x}-\mathbf{x}')
\te
where $S^{\mu\nu}_{\alpha\beta}$ is the transverse and traceless projector \eqref{Transverse and traceless projectors}. Fourier transforming this expression we get
\be
\langle \zeta_{\mathbf{k},\sigma}^{\rm ini}\zeta_{\mathbf{k}',\sigma'}^{\rm ini}\rangle=(2\pi)^3  \mathcal{P}_\zeta(k) \delta_{\sigma\sigma'}\delta(\mathbf{k}+\mathbf{k}')
\te
where $P_\zeta(k)$ is the flat spectrum of initial fluctuations
\be
\mathcal{P}_\zeta(k)=\frac{75}{8}\frac{1}{a^4_{-1}}.\label{initial spectrum zeta}
\te
Note that, up to a numerical factor, this spectrum is inversely proportional to the viscous contribution to the energy-momentum tensor of the fluid \eqref{Perturbed fluid EMT}.

\subsection{Fluid and gravitational waves energy spectrum}
\label{Fluid and GW energy spectra}

In our model both the fluid fluctuations $\zeta^{\mu\nu}$ and  $\xi_{\mu\nu\rho}$, and the gravitational wave amplitudes $h_{\mu\nu}$ are stochastic fields with zero mean.  However, they contribute to the energy, computed to quadratic order in the fields. We thus define the fluid energy from the expectation value of  $T_f^{00}$ component of the fluid energy–momentum tensor \eqref{Fluid EMT}, keeping terms up to second order in the nonequilibrium tensors,
\be
\label{Fluid perturbations energy density}
\delta \rho_{f} 
= \frac{1}{15} a^4_0 \,\langle \zeta_{\mu\nu}\zeta^{\mu\nu} \rangle 
+ \frac{3}{105} a^6_{-2}\,\langle \xi_{\mu\nu\rho}\xi^{\mu\nu\rho} \rangle\,.
\te
We are interested in the spectrum of energy density per logarithmic wave-number interval normalized by the critical density $\rho_c$, defined as
\be
\Omega^{f}_k(\theta)=\frac1{\rho_c}\,\frac{d\,\delta \rho_f}{d\ln k}.
\te
where the critical density is the mean value of the inflaton energy density $\rho_c\approx \langle\rho_\phi\rangle=\frac{3}{8}\,\Omega^2 \phi_0^2$.

We make explicit the contribution from the transfer functions $\tilde\zeta_\kappa(\theta)$ and $\tilde\xi_\kappa(\theta)$ from eq. (\ref{Stochastic+deterministic Fourier modes}), and the initial spectrum $\mathcal{P}_\zeta(k)$ in Eq. \eqref{initial spectrum zeta}. We get
\be
\label{Fluid energy fraction}
\Omega^f_\kappa(\theta)=\frac{15M\Omega}{18\pi^2\phi_0^2}\frac{L_{4,0}}{L_{4,-1}}\kappa^3\lc |\tilde{\zeta}_{\kappa}(\theta)|^2+\frac{3\kappa^2}{49}\frac{L_{6,-2}}{L_{4,0}}|\tilde{\xi}_{\kappa}(\theta)|^2 \rc
\te
where $\tilde\zeta_\kappa$ is obtained by solving \eqref{eq:y-eq} for $y_\kappa$ and using \eqref{Dissipative term substitution}, while $\tilde{\xi}_\kappa$ is obtained by solving Eq.~\eqref{Perturbation equations simplified 1} and expressed as a function of $\tilde{\zeta}_\kappa$. It can be seen, from the behaviour of the $L_{k,l}(z)$ functions (see Appendix \ref{Section_L functions}), that the contribution of $\tilde\xi_k$ to the fluid energy density is subdominant.

On the other hand, the energy density of the gravitational waves is given by \cite{1973_MTW,2008_Maggiore}
\be
\label{Definition GW energy}
\rho_{\rm GW}(\theta) = \frac{1}{4} m_p^2 
\langle \dot h_{ij}(\mathbf{x},t)\,\dot h_{ij}(\mathbf{x},t) \rangle
= m_p^2 \Omega^2 \int \frac{d^3k}{(2\pi)^3} 
\mathcal{P}_\zeta(k)\,|\dot g_k(t)|^2\,.
\te
Then, the gravitational waves energy density fraction is
\be
\label{GW enrgy fraction_0}
\Omega^{\text{GW}}_\kappa(\theta)=\frac{25}{2\pi^2}\frac{m_p^2\Omega^5}{\phi_0^2 a^4_{-1}}\kappa^3| g'_\kappa(\theta)|^2.
\te

Let us analyze these expressions. As we have already mentioned, due to its coupling to the coherently oscillating condensate, the fluid undergoes parametric amplification. This leads to the amplification of certain momentum bands, with the peak of the resonance band growing exponentially in time as $y_\kappa \sim e^{\tilde\mu_{\text{max}}\theta}$, where $\tilde\mu_\text{max}$ denotes the maximum Floquet exponent (the peak of the resonance band). As a rough estimate, the energy fraction of the fluid \eqref{Fluid energy fraction} evolves as
\be
\label{Fluid energy density evolution}
\Omega^f_\kappa(\theta)\approx \frac{15}{18\pi^2}\frac{M\Omega}{\phi_0^2}  e^{2\mu_\text{max}\theta}
\te
where we defined $\mu_{\text{max}}=\tilde\mu_\text{max}-\tilde\tau^{-1}$. Note that only if the amplification of $y_\kappa$ exceeds the dissipation, characterized by the relaxation time $\tilde\tau$, will the nonequilibrium fluctuations of the fluid ($\tilde\zeta_\kappa$) grow, which requires $\mu_\text{max}=\tilde\mu_{\text{max}}-\tilde\tau^{-1}>0$. In this case, we can estimate the time $\theta_{\text{nl}}$ at which nonlinear effects become relevant by solving $\Omega^f_\kappa(\theta_{\text{nl}})=1$, which gives $\theta_{\text{nl}}\sim \ln(\phi_0^2/M\Omega\kappa_*^3)/2\mu_\text{max}$, where $\kappa_*$ is the most amplified momenta (the peak of the spectrum). Our results are valid as long as $\theta\leq\theta_{\text{nl}}$. This allows us to estimate the energy fraction of the gravitational waves when nonlinear effects become relevant
\be
\Omega^{\text{GW}}_\kappa(\theta_{\text{nl}})\approx \frac{8n_*}{15\pi^2}\frac{M^4}{\Omega^2 m_p^2}\frac{\mu_\text{max}^2}{\kappa_*^4} L_{4,-1}.
\te
We see that, although the time evolution of both energy densities is similar, the energy fraction of the gravitational waves is several orders of magnitude smaller than that of the fluid. This suppression arises from the effective coupling between the fluid and the gravitational waves, scaling as $u/\Omega^3$, and through the function $L_{4,-1}$, which remains small for all relevant values of $z=M/T$ considered in this work.

\section{Evolution of the tensor modes}
\label{Section_Evolution of the tensor modes}
In this section we describe the evolution of the tensor modes of both the fluid and the gravitational waves by numerically solving the two coupled dynamical equations, namely \eqref{eq:y-eq} and \eqref{Einstein equation for GW (final)}, whose structure is similar to the one in the standard theory of reheating \cite{2019_Lozanov}. In the first subsection we obtain the fluid energy fraction, and in the second subsection we study the evolution of the energy fraction of the gravitational waves.

\subsection{Parametric amplification of the fluid}

We now study the evolution of the variable $y_\kappa$, by solving eq. \eqref{eq:y-eq} with the time dependent frequency eq. \eqref{Mathieu-like equation frequency}. We find that this variable is amplified through a narrow resonance parametric amplification scenario.

It is interesting to compare this effect to the behavior of the first-order fluctuations of the fields in the standard theory of Reheating \cite{2019_Lozanov}. In the standard reheating scenario the Mathieu equation takes the form $\omega^2 = A + 2\delta \cos{2\theta}$, where $A \geq 0$. In our model, we have $A = \kappa^2/7r - \delta^2/2$. 

Because the instantaneous frequency may be imaginary, we may expect there will be a process of spinodal decomposition \cite{1989_Calzetta,2001_Felder,2001b_Felder,2002_Copeland} along with the parametric amplification. However this effect occurs only for long wavelength modes and it is not relevant for the production of gravitational waves in this model.  

Another difference with the usual models of reheating is the presence of two frequency components: the lower frequency $2\Omega t$, and the higher frequency $4\Omega t$, whose contribution is of higher order in $\delta$. We evaluate the Floquet exponent $\tilde\mu_\kappa$ for the variable $y_\kappa$,  Using standard perturbative methods \cite{1976_Landau}, valid since $\delta$ is a small parameter, we obtain
\be
\label{Analytical Floquet exponent}
\tilde\mu_\kappa=\frac{\delta}{2}\sqrt{1-\left[ \frac{1}{\delta}\lc\frac{\kappa^2}{7r}-1-\frac{3}{8}\delta^2 \rc\right]^2}.
\te
We see that the high-frequency component induces a small correction to the usual result.

\begin{figure}
    \centering
    \includegraphics[width=\linewidth]{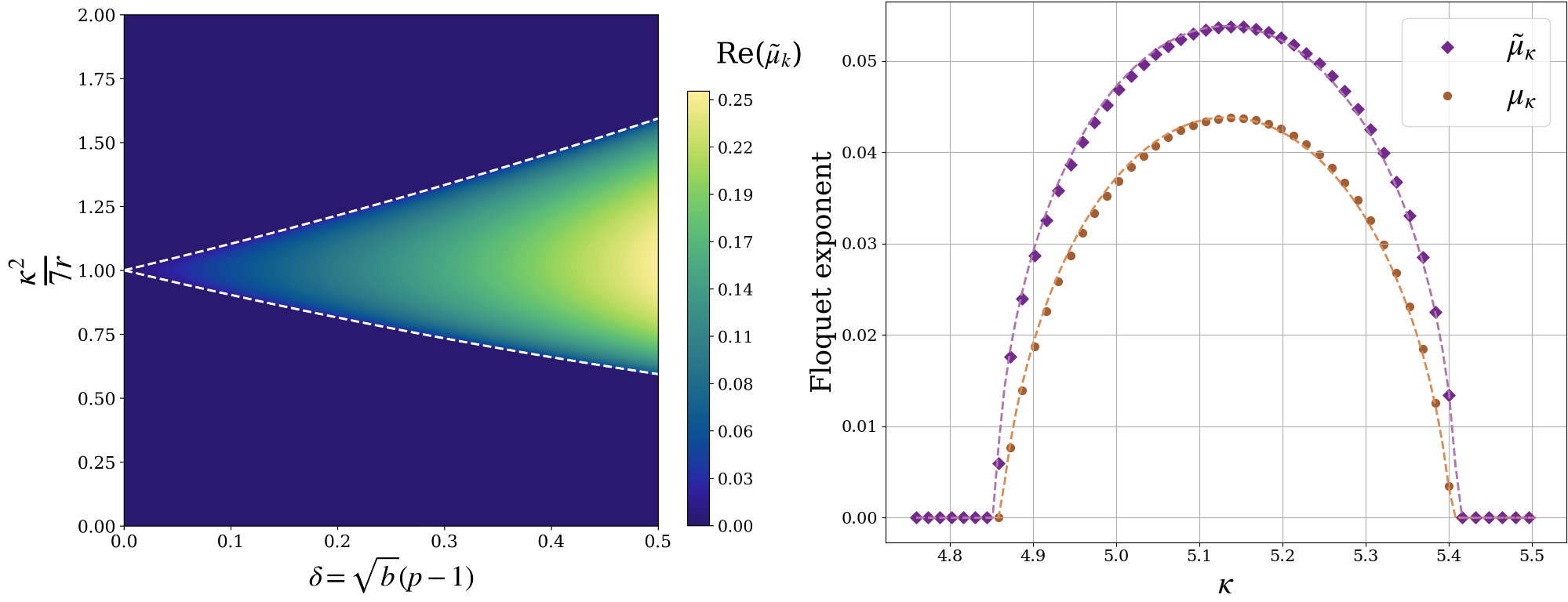}
    \caption{(Left) Instability regions and Floquet exponents of the Mathieu-like equation \eqref{Mathieu-like equation} for $y_\kappa$. The resonance band is centered at $\kappa^2=7r$, with maximum amplification at $\delta=1/2$. White dashed lines mark the instability boundaries from \eqref{Analytical Floquet exponent}, in excellent agreement with the numerical results. (Right) Numerical (dashed lines) and analytical (points) Floquet exponents $\tilde{\mu}_\kappa$ and $\mu_\kappa$ for $y_\kappa$ and $\zeta_\kappa$, respectively. The parameters used were: $\delta=0.11$, $z=20$, $q=1$, $\tilde\tau=85$.} 
    \label{fig:Instability chart + Floquet exponents}
\end{figure}

In the left panel of Fig. \ref{fig:Instability chart + Floquet exponents} we show the instability chart featuring the real part of the Floquet exponent $\tilde\mu_\kappa$ as a function of $\kappa^2/7r$ and $\delta=\sqrt{b}(p-1)$. The white lines indicate the limits of the resonant band predicted by \eqref{Analytical Floquet exponent}, which are in excellent agreement with the numerical computation. For a given value of $\delta$ the maximum amplification with a Floquet exponent $\tilde\mu_{\text{max}}=\delta/2$ occurs for momentum $\kappa_*$ such that
\be
\kappa_*^2=7r\lc 1+\frac38\delta^2 \rc\,.
\te

\begin{figure}[b!]
    \centering
    \includegraphics[width=.68\linewidth]{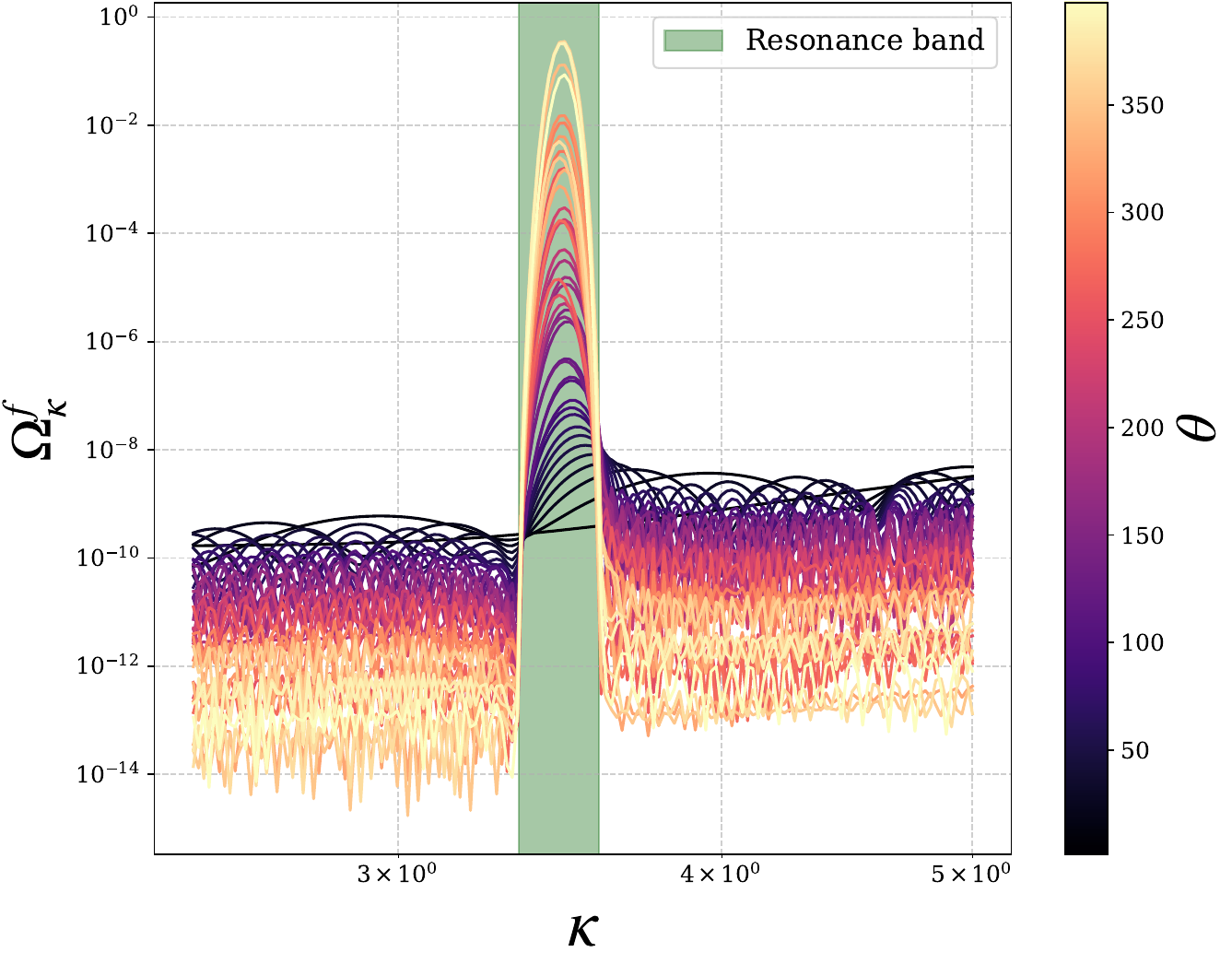}
    \caption{Evolution of the fluid energy fraction spectrum for several times, up to $\theta_{\text{nl}}\approx400$. We see that there is an exponential amplification of the fluid in the resonance band (highlighted in green in the background) and that modes outside it decay exponentially due to the usual dissipation in a theory with relativistic real fluids.} 
    \label{fig:Fluid energy evolution}
\end{figure}

However, the Floquet exponent of physical relevance is $\mu_\kappa$ the one associated to the variable $\tilde\zeta_\kappa$, instead of $y_\kappa$. This exponent is obtained by subtracting the non-oscillatory contribution from the transformation that relates $y_\kappa$ to $\tilde\zeta_\kappa$, namely \eqref{Dissipative term substitution}. Therefore, the Floquet exponent of the variable $\tilde\zeta_\kappa$ is given by
\be
\label{Floquet exponent zeta tensor}
\mu_\kappa=\tilde\mu_\kappa-\frac{1}{\tilde\tau}.
\te
We see that this exponent is bounded by $\mu_{\text{max}}\leq 0.25$, which is in close agreement with the maximum exponent in the standard theory of reheating—when the expansion of the universe is not considered—namely, $\mu_{\text{max}}\leq 0.28$ \cite{1997_KLS}. In the right panel of Fig. \ref{fig:Instability chart + Floquet exponents}, we show the numerical results for the Floquet exponents $\tilde{\mu}_\kappa$ and $\mu_\kappa$, plotted as purple and brown lines, respectively. The analytical results from \eqref{Analytical Floquet exponent} and \eqref{Floquet exponent zeta tensor} are represented by points in the corresponding colors. We observe an excellent agreement between the numerical and analytical results.

To go beyond this perturbative arguments we solve numerically the evolution of $y_\kappa$ in a range of $\kappa$. We use the numerical solution to evaluate the fluid energy fraction \eqref{Fluid energy fraction} and plot it as a function of $\kappa$ and for several times as shown in Fig. \ref{fig:Fluid energy evolution}. In this numerical simulation, the parameters used were $\lambda=10^{-14}$, $z=5.8$, $\phi_0 = m_p$, $\tilde\tau \approx 110$, and $q = 1.5$, corresponding to a resonance parameter of $\delta = 0.7$. We stop the numerical solution at the time $\theta_{\text{nl}}\approx 400$ when the energy density in the peak reaches the critical density. We observe good agreement between the numerical solution and the analytic approximation \eqref{Fluid energy density evolution} derived in section \ref{Fluid and GW energy spectra}, up to this time. The analytically predicted resonance band is highlighted in green in the background; the peak of this resonance band is in the momenta $\kappa_*\approx3.46$.

We observe that the resonance band is amplified, while the overall behavior of the fluid energy density follows a pattern similar to that of the standard theory of reheating \cite{2019_Lozanov}. Modes outside the resonance band decay due to the dissipative effects of this hydrodynamic theory, since for these modes $\tilde{\zeta}_k \sim e^{-\theta / \tilde{\tau}}$.

\subsection{Gravitational wave production}

\begin{figure}[b]
    \centering
    \includegraphics[width=.68\linewidth]{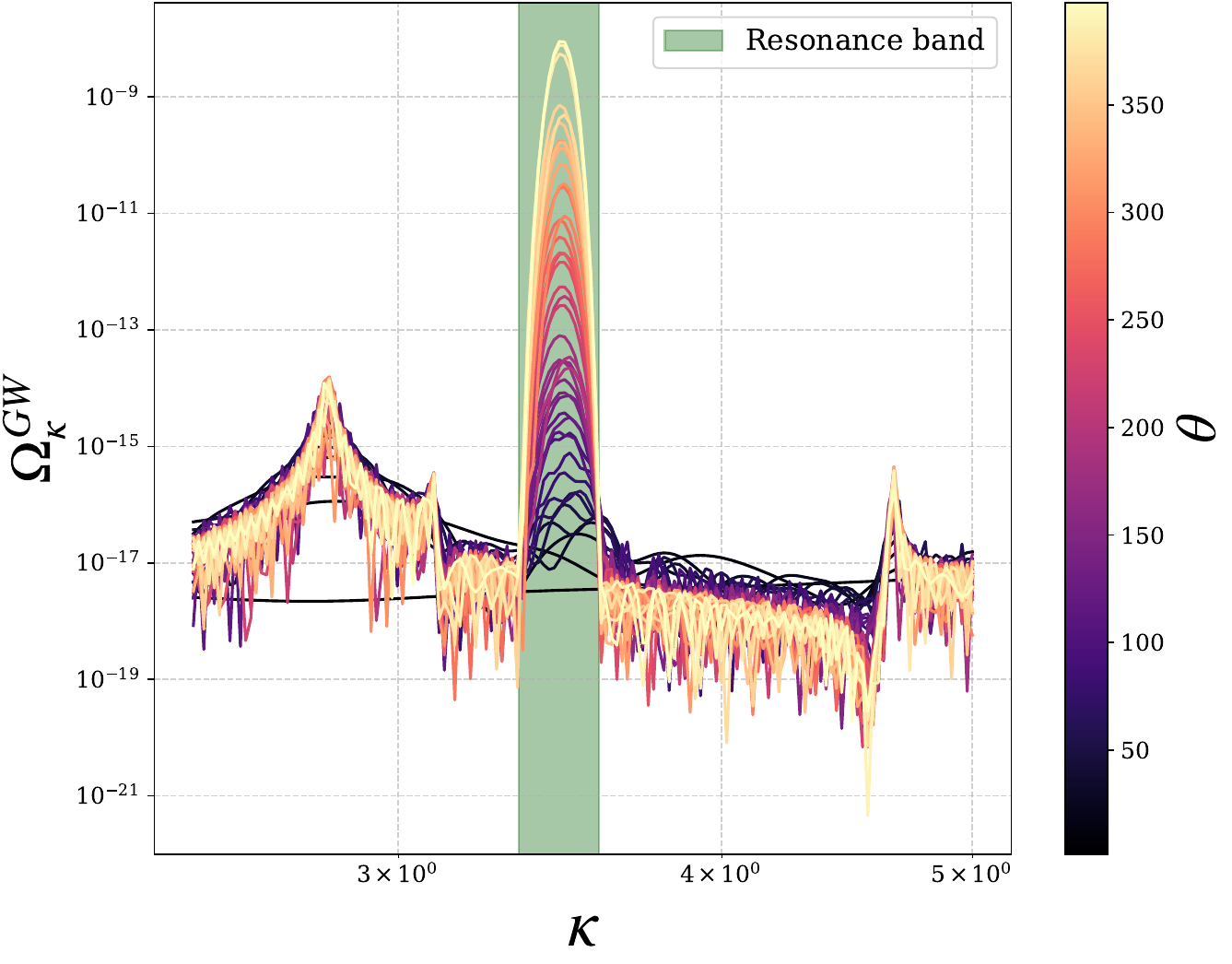}
    \caption{Evolution of the gravitational-wave energy fraction spectrum at several times, up to $\theta_{\text{nl}}\approx400$. The main amplification occurs within the parametric resonance band (highlighted in green), while outside this region a rapid growth due to standard resonance takes place, which quickly saturates.} 
    \label{fig:GW energy evolution}
\end{figure}

In this subsection we study the production of gravitational waves sourced by the fluid tensor modes. We numerically solve \eqref{Einstein equation for GW (final)} which allows us to get the energy fraction of the gravitational waves \eqref{GW enrgy fraction_0}. In Fig. \ref{fig:GW energy evolution}, in analogy with what we have done with the fluid energy fraction, we plot this energy fraction as a function of $\kappa$ and for several times.

We observe that the amplification of the gravitational wave energy fraction takes place within the same resonance band as the fluid, as expected when the source of gravitational waves has a well-defined peak in momentum space \cite{2018_Caprini}. Outside the resonance band, the driving of the gravitational field by the fluid makes the gravitational wave amplitudes to grow rapidly and subsequently stabilize at relatively short times, $\theta \sim 70$. 

At time $\theta_{\text{nl}}\approx 400$ when our numerical solution stops, the GW energy fraction at the peak $\kappa_*$ reaches $\Omega^{\text{GW}}_{\kappa_*}(\theta_{\text{nl}})=9\cdot10^{-9}$, much larger than the analytical prediction, $\Omega^{\text{GW}}_{\kappa*}(\theta_{\text{nl}})\approx 10^{-12}$. This may be understood as the effect of oscillations in the fluid modes, which are small with respect to the exponentially growing peak, but still resonant with the gravitational fluctuations.

\begin{figure}
   \centering
    \includegraphics[width=.68\linewidth]{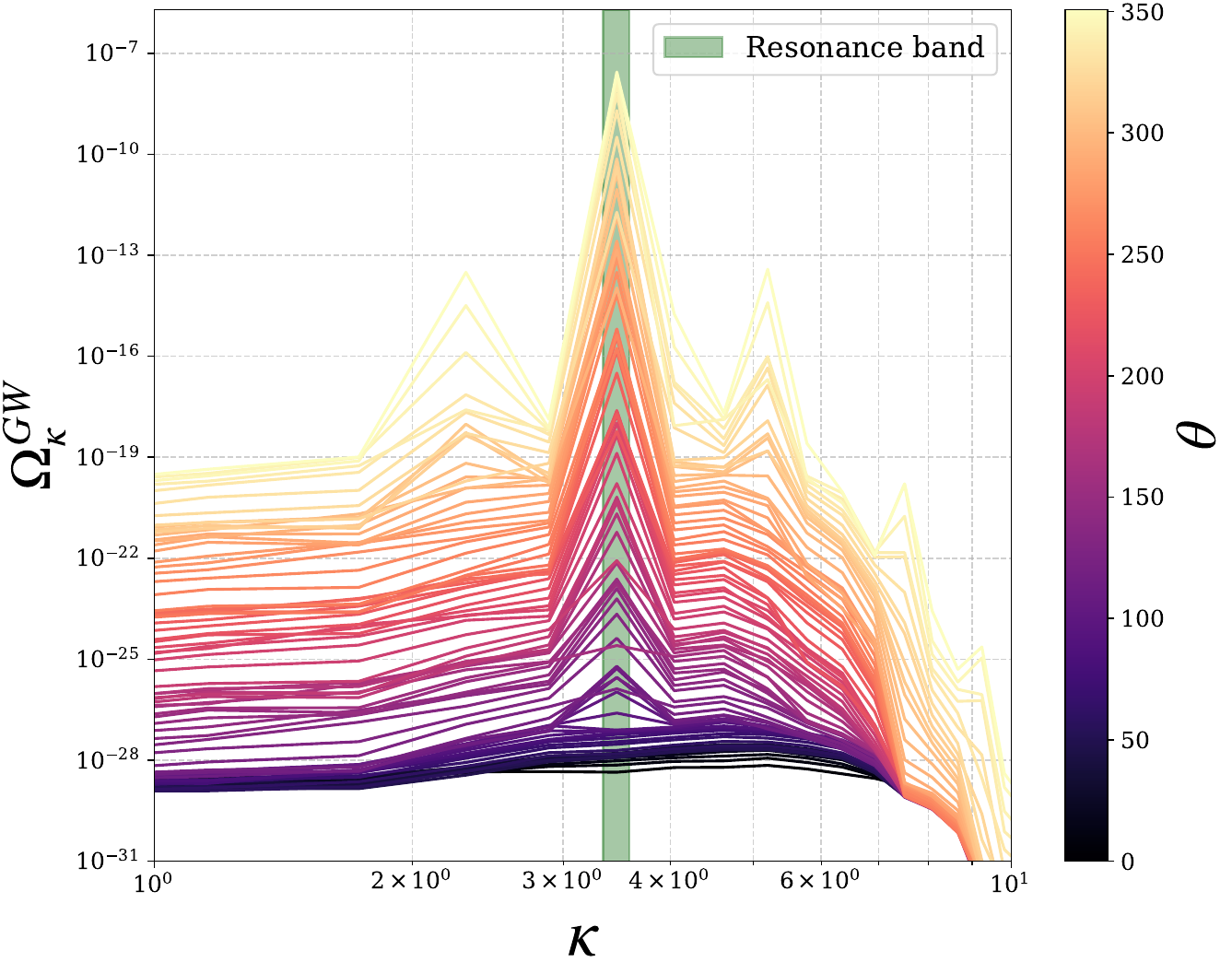}
   \caption{Gravitational-wave energy fraction spectrum obtained with CosmoLattice for the model \(V(\phi)=\lambda\phi^4/4\) with \(\lambda=10^{-14}\) and initial field \(\phi_0=2m_p\). The position of the resonance band (highlighted in green) coincides with that found in the hydrodynamical model, and the peak of the spectrum appears at the same location, although CosmoLattice yields a broader distribution.} 
  \label{fig:CosmoLattice GW energy evolution}
\end{figure}

It is interesting to compare the results of our phenomenological model with a widely used code such as Cosmolattice \cite{2020_Figueroa,2023_Figueroa}. In Fig. \ref{fig:CosmoLattice GW energy evolution} we plot the spectrum of gravitational waves in a model with inflation potential $V(\phi)=\lambda\phi^4/4$ with $\lambda=10^{-14}$, which is taken as the same value with the hydrodynamical theory, and consider no coupling with other scalar fields. The run the simulation with initial field $\phi_0=2m_p$, which should give a similar result as in the hydrodynamical model. We see that the results in Figs. \ref{fig:GW energy evolution} and \ref{fig:CosmoLattice GW energy evolution} are in good agreement, specially regarding the position of the peak, Cosmolattice predicting a broader peak. As the position of the peak is the most important information to be extracted from the simulation, we find this result encouraging.

\section{Conclusions}
\label{Section_Conclusions}

In this work, we have developed a causal hydrodynamic model aimed at providing an effective macroscopic description of the field-theoretic dynamics during the early stages of reheating. In particular, we describe the energy transfer from the oscillating inflaton condensate to a relativistic fluid, which effectively represents the condensate fluctuations, using the framework of a divergence-type hydrodynamic theory \cite{geroch_1990,1986_Liu,2020_Cantarutti_Calzetta}, derived from kinetic-theory considerations. This approach captures, in an effective way, the essential features of the underlying microscopic field interactions, such as dissipation, parametric amplification, and mode coupling, within a stable and causal formulation that includes propagating tensor degrees of freedom.

Our results show that the coupling between the oscillating condensate and the fluid leads to a parametric resonance in the tensor sector of the fluid. The viscous stress tensor undergoes exponential amplification in a well-defined momentum range, sourcing the production of gravitational waves with a characteristic spectral peak. The position of this peak, and the early-time growth of the corresponding energy density, are in good agreement with results obtained from full numerical lattice simulations performed with the public code CosmoLattice \cite{2020_Figueroa,2023_Figueroa}, while requiring considerably less computational effort.

The hydrodynamic framework introduced here thus provides an effective and physically transparent tool to explore the reheating process in a way complementary to first-principles field-theoretic simulations. It also offers a natural setting to incorporate dissipative effects, stochastic fluctuations, and nonequilibrium dynamics in a consistent relativistic formulation.

Future work will focus on extending this approach to include scalar and vector perturbations, a realistic mass spectrum for the fluid, gauge-field interactions, and the backreaction of gravitational waves. Ultimately, a complete hydrodynamic treatment of reheating could serve as a bridge between microscopic particle dynamics and macroscopic cosmological observables, offering a complementary path toward understanding the generation of stochastic gravitational waves in the early Universe. Moreover, this hydrodynamic scheme could also prove useful for describing the turbulent stages of reheating, as well as the thermalization process itself \cite{2002_Grana,Calzetta_2021_prd,2003_Micha,2004_Micha}.

\begin{acknowledgments}

We acknowledge financial support from Universidad de Buenos Aires through Grant No. UBACYT~20020170100129\allowbreak BA, CONICET Grant No.~PIP2017/19:\allowbreak 11220170100817\allowbreak CO and ANPCyT Grant No.~PICT~2018:\allowbreak 03684.

\end{acknowledgments}

\appendix

\section{Derivation of the hydrodynamical equations}
\label{Section_Derivation of Hydro equations}

In this Appendix we derive the hydrodynamical equations of the model \eqref{Conservation laws}. We consider a relativistic causal fluid, where dissipative dynamics are treated by including auxiliary degrees of freedom \cite{2013_RezzollaZanotti}. 

Such theories can be derived as the low frequency, long wavelength limit of a quantum field theory, where the field excitations are described as quasi-particles with a temperature dependent mass \cite{calzetta1988, jeon1993, jeon1995, jeon1996, calzetta2000, 2008_Calzetta_Hu}.

The gas of quasiparticles obeys a Boltzmann equation
\be
\label{Boltzmann equation}
p^\nu \nabla_\nu f - I_{\text{col}} = - F^\nu \frac{\partial f}{\partial p^\nu}.
\te
The external force is $F_\mu = -\tfrac{1}{2} \partial_\mu M^2$, where $M$ the quasi-particle mass, which depends both on the temperature of the quasi-particles and the amplitude of the condensate. Because of particle-antiparticle symmetry we assume the gas has vanishing chemical potential. 

We shall further reduce the kinetic description to a hydrodynamic one by applying the method of moments \cite{1986_Liu}. We first choose a preferred set of functions $c_A$, and then compute their moments 
\be 
C^{\mu}_{A}=\int Dp\; p^{\mu}c_{A}f,
\label{moments}
\te 
where the invariant measure $Dp$ is defined in eq. (\ref{Invariant measure}). The $C^{\mu}_{A}$ are regarded as the fundamental observables of the theory. Because the energy-momentum tensor plays a special role, it is natural to choose the first function $c_0 =p_{\nu}$. Then, further moments are added in order to obtain a causal and stable theory \cite{struchtrup2003,denicol2012,jaiswal2013,brito2022,brito2023,panday2024,jeon2024,jeon2025}.

Once the $c_{A}$ functions are chosen, to obtain a close system of equations the 1pdf is approximated by the maximum entropy distribution function, subject to equations (\ref{moments}) as so many constraints
\be 
f\approx e^{\sum_{A}\lambda_{A}c_{A}}
\te 
whereby a set of Lagrange multipliers $\lambda_{A}$ are introduced. Given our choice of $c_0$ it is natural to choose $\lambda_0=\beta_{\nu}$, which is further identified with the inverse temperature four vector. The Lagrange multipliers become the dynamical degrees of freedom of the theory, since the $C^\mu_A$ moments are functions of them.

We obtain the dynamical equations by taking moments of the kinetic equation 
\be
\label{moments Boltzmann equation}
\int Dp\; c_A\left[ p^\nu \nabla_\nu f - I_{\text{col}} + F^\nu \frac{\partial f}{\partial p^\nu}\right]=0.
\te
These equations imply positive entropy production \cite{2020_Cantarutti_Calzetta}. 

Equations (\ref{moments Boltzmann equation}) define the divergence of the moments \eqref{moments} as linear combinations of derivatives of the Lagrange multipliers, multiplied by suitable transport coefficients. These coefficients may be derived by fitting the hydrodynamic response functions to the low frequency, long wavelength limit of the underlying quantum field theory \cite{2021_Perna_Calzetta,jeon2025}. 

Since we assume that the collision term takes the relaxation-time-approximation form (see Eq.~(\ref{collision integral}) below), characterized by a single relaxation time~$\tau$, all transport coefficients are proportional to it. However, the theory can be straightforwardly generalized to allow for multiple relaxation times \cite{florkowski2016,huang2015}. We discuss the relation between the relaxation time and the shear-viscosity transport coefficient in Appendix~(\ref{Section_Relaxation time calculation}); analogous relations for the other transport coefficients can be derived following a procedure similar to that of \cite{brito2023,jeon2025}.

In this work our choice of the preferred moments shall be 
\bea 
c_0&=p^{\mu}\nn
c_1&=S^{\alpha\beta}_{\mu\nu}\frac{p^\mu p^\nu}{(-u_\lambda p^\lambda)}
\tea 
and 
\be 
c_2=S^{\alpha\beta\gamma}_{\mu\nu\rho}\frac{p^\mu p^\nu p^\rho}{(-u_\lambda p^\lambda)^2}
\te 
where the projectors $S^{\alpha\beta}_{\mu\nu}$ and $S^{\alpha\beta\gamma}_{\mu\nu\rho}$ are defined in eq. (\ref{Transverse and traceless projectors}). Thereby we must include the Lagrange multipliers
\bea 
\lambda_0&=\beta_{\mu}\nn 
\lambda_1&=\zeta_{\mu\nu}\nn 
\lambda_2&=\xi_{\mu\nu\rho}
\tea 
where $\zeta_{\mu\nu}$ and $\xi_{\mu\nu\rho}$ are totally symmetric, transverse and traceless. The powers of $(-u_\lambda p^\lambda)$ appearing in $c_1$ and $c_2$ are introduced because, in test cases where kinetic theory and hydrodynamic calculations can be performed simultaneously and directly compared, they yield significantly improved agreement between the two formalisms \cite{2020_Cantarutti_Calzetta,2021_Perna_Calzetta,2022_Calzetta}. As a consequence, our final equations differ from those presented in \cite{panday2024,brito2023,jeon2024,jeon2025}. Nevertheless, the different sets of equations are easily seen to describe the same physics.

In summary, we parametrize the 1pdf as
\be
\label{1pdf}
f[x^\mu,p^\mu,\beta_\mu,\zeta_{\mu\nu},\xi_{\mu\nu\rho}]
=\exp(\beta_\mu p^\mu
+\zeta_{\mu\nu}\frac{p^\mu p^\nu}{(-u_\lambda p^\lambda)}
+\xi_{\mu\nu\rho}\frac{p^\mu p^\nu p^\rho}{(-u_\lambda p^\lambda)^2})
\te
where $\beta_\mu=\frac{u_\mu}{T}$, with $T$ the temperature of the fluid and $u_\mu$ its four-velocity, satisfying $u_\mu u^\mu=-1$. $\zeta_{\mu\nu}$ and $\xi_{\mu\nu\rho}$ are two non-equilibrium tensors. Both non-equilibrium tensors are totally symmetric, traceless and transverse.

The equations of motion for the fluid are obtained by taking moments of the Boltzmann equation according to
\bea
\label{Moments of the Boltzmann equation}
\int Dp\, p^\mu\left[p^\tau f_{;\tau}-I_{col}\right]&=-F^\tau\int Dp\, p^\mu\frac{\partial}{\partial p^\tau}f\\
\int Dp\, S^{\alpha\beta}_{\mu\nu}\frac{p^\mu p^\nu}{(-u_\lambda p^\lambda)} \left[p^\tau f_{;\tau}-I_{col}\right]&=-F^\tau\int Dp\, S^{\alpha\beta}_{\mu\nu}\frac{p^\mu p^\nu}{(-u_\lambda p^\lambda)} \frac{\partial}{\partial p^\tau}f\\
\int Dp\, S^{\alpha\beta\gamma}_{\mu\nu\rho}\frac{p^\mu p^\nu p^\rho}{(-u_\lambda p^\lambda)^2} \left[p^\tau f_{;\tau}-I_{col}\right]&=-F^\tau\int Dp\, S^{\alpha\beta\gamma}_{\mu\nu\rho}\frac{p^\mu p^\nu p^\rho}{(-u_\lambda p^\lambda)^2} \frac{\partial}{\partial p^\tau}f.
\tea
where $Dp$ is the invariant measure in momentum space \eqref{Invariant measure}. Note that, in principle, when integrating these equations to obtain eq.~\eqref{Conservation laws}, one must also take into account derivatives acting on the integration measure. These contributions are given by
\be
\int \frac{d^4p}{(2\pi)^3} \, c_A f
\left[ p^\nu \nabla_\nu + F^\nu \frac{\partial}{\partial p^\nu} \right]
\delta(p^2+M^2)
=
\int \frac{d^4p}{(2\pi)^3} \, c_A p^\nu f
\left( M^2_{;\nu} + 2F_\nu \right)
\delta'(p^2+M^2),
\te
where $\delta'$ denotes the derivative of the Dirac delta function with respect to its argument. These terms vanish identically due to the specific choice of the external force.

These equations must be projected with the tensors defined in \eqref{Transverse and traceless projectors} because not all components of the non-equilibrium tensors are independent, and hence conservation equations are imposed only on their transverse and traceless parts. Upon integration by parts, these equations (\ref{Moments of the Boltzmann equation}) take the form of conservation laws \eqref{Conservation laws} with the constitutive relations
\bea
\label{Tensors in the conservation laws}
M_T^2&=\int Dp\,f\\
B^{\mu_1...\mu_n}&=\int Dp\;\frac{p^{\mu_1}...p^{\mu_n}}{(-u_\lambda p^\lambda)^n}f\\
A^{\mu_1...\mu_n}&=\int Dp\;\frac{p^{\mu_1}...p^{\mu_n}}{(-u_\lambda p^\lambda)^{n-2}}f\\
I^{\mu_1...\mu_n}&=\int Dp\;\frac{p^{\mu_1}...p^{\mu_n}}{(-u_\lambda p^\lambda)^{n-1}}I_{\text{col}}\\
\tea
with $M_T$ the thermal mass. In particular, the energy-momentum tensor of the fluid $T^{\mu\nu}_f$ is given by
\be
\label{Fluid EMT}
T^{\mu\nu}_f=\int Dp\,p^\mu p^\nu f.
\te
For the collision integral we employ the Anderson--Witting relaxation-time approximation \cite{1974_Anderson,1974b_Anderson},
\be
\label{collision integral}
I_{\text{col}} = \frac{1}{\tau} \, u_\mu p^\mu \left[f - f_0\right],
\te
where $f_0$ is the equilibrium function given by \eqref{local_equilibrium_distribution} and $\tau$ is the relaxation time derived in Appendix~\ref{Section_Relaxation time calculation}.

Since the collision term obeys an $H$ theorem, these equations enforce the Second Law of thermodynamics \cite{2020_Cantarutti_Calzetta}, Eq. \eqref{entropy_production}, with entropy flux given by
\be 
S^{\mu}=\int Dp\; p^{\mu}\;f[1-\ln f].
\label{entropyflux}
\te

We consider small perturbations around equilibrium and expand the distribution function \eqref{1pdf} up to first order in $\zeta_{\mu\nu}$ and $\xi_{\mu\nu\rho}$ according to
\bea
f\simeq f_0\lc 1+\zeta_{\mu\nu}\frac{p^\mu p^\nu}{(-u_\lambda p^\lambda)}
+\xi_{\mu\nu\rho}\frac{p^\mu p^\nu p^\rho}{(-u_\lambda p^\lambda)^2} \rc \,.
\tea
Thus, up to first order, the tensors in \eqref{Tensors in the conservation laws} that appear in the conservation laws \eqref{Conservation laws} are
\bea
\label{Tensors in the conservation laws first order}
A^{\mu\nu\rho}&=T_1^{\mu\nu\rho}+2 C^5_2(2) u^\mu\zeta^{\nu\rho}+6 C^6_3(3) \xi^{\mu\nu\rho}\\
A^{\mu\nu\rho\sigma}&=T_2^{\mu\nu\rho\sigma}+2 C^6_3(2) \left\{ u^\mu u^\nu\zeta^{\rho\sigma}\right\}+2C^6_3(3)\left\{ \Delta^{\mu\nu}\zeta^{\rho\sigma} \right\}+6 C^7_4(3)\left\{ u^\mu\xi^{\nu\rho\sigma}\right\}\\
A^{\mu\nu\rho\sigma\lambda}&=T_3^{\mu\nu\rho\sigma\lambda}
+2C^7_4(2)\{u^\mu u^\nu u^\rho \zeta^{\sigma\lambda}\}
+2C^7_4(3)\{ u^\mu \Delta^{\nu\rho}\zeta^{\sigma\lambda}\}\\
&+6 C^8_5(3)\{u^\mu u^\nu \xi^{\rho\sigma\lambda}\}+6 C^8_5(4)\{\Delta^{\mu\nu} \xi^{\rho\sigma\lambda}\}\\
I^{\mu\nu}&=\frac{2}{\tau}\,C^4_1(2)\,\zeta^{\mu\nu}\\
I^{\mu\nu\rho}&=\frac{6}{\tau}\,C^6_3(3)\,\xi^{\mu\nu\rho}\\
B^\mu&=T_1^\mu\\
B^{\mu\nu}&=T_2^{\mu\nu}+2C^4_3(2)\zeta^{\mu\nu}\\
B^{\mu\nu\rho}&=T_3^{\mu\nu\rho}+2C^5_4(2)\{u^\mu \zeta^{\nu\rho}\}+6C^6_5(3)\xi^{\mu\nu\rho}
\tea
where the tensors appearing inside the braces are understood to be completely symmetrized. Here we defined
\be
T_m^{\mu_1...\mu_n}=\int Dp\,\frac{p^{\mu_1}\ldots p^{\mu_n}}{(-u_\lambda p^\lambda)^m}f_0.
\te
These tensors, after integration, can be written in terms of the fluid four-velocity and the coefficients $C^n_m(k)=a^{2k}_{n-m-2k}/(2k-1)!!$, as in \eqref{General tensors structure}. The functions $a^{2k}_{n-m-2k}$ were defined in eq. \eqref{a^k_l functions}. Then, the first-order dynamical equations take the form \eqref{Dynamical equations fluid}. 
This scheme allows us to write the equations for the background and the tensor sector of the fluid fluctuations as
\bea
\label{Appendix_Full_Hydro_Eqs}
a^0_{2,\nu}\; u^\mu u^\nu + \frac{1}{3} a^2_{0,\nu}\; \Delta^{\mu\nu}+\left(a^0_{2}+\frac13a^2_{0}\right)( u^\mu_{;\nu} u^\nu + u^\mu u^\nu_{;\nu})+\frac{2}{15}\nabla_\nu \left( a^4_{-1}\;\zeta^{\mu\nu}\right)&=F^\mu M_T^2\\
\left(5a_0^2 -a^4_{-2}\right)\lc \Delta^{(\alpha}_{\nu}u^{\beta);\nu}-\frac13 u^\mu_{;\mu}\Delta^{\alpha\beta}\rc+a^4_{-1}\;\left[\dot\zeta^{\alpha\beta} +\zeta^{\alpha\beta}\left( \frac{1}{\tau}+\frac{\Dot{a}^4_{-1}}{a^4_{-1}} \right)\right]+ \frac{3}{7}a^6_{-3}\xi^{\alpha\beta\rho}_{,\rho}
&=a^4_{-3}\;\zeta^{\alpha\beta}F^\tau u_\tau\\
a^6_{-3}\left[ \zeta^{(\alpha\beta}_{;\sigma}\Delta^{\gamma)\sigma}-\frac25 \Delta^{(\alpha\beta}\zeta^{\gamma)\sigma}_{;\sigma}\right]
+a^6_{-3}\left[\Dot{\xi^{\alpha\beta\gamma}}+ \xi^{\alpha\beta\gamma}\left( \frac{1}{\tau}+\frac{\Dot{a}^6_{-3}}{a^6_{-3}} \right)\right]
&=2a^6_{-5}\xi^{\alpha\beta\gamma}F^\sigma u_\sigma
\tea
where terms such as $\Delta^{(\alpha}_{\nu} u^{\beta);\nu}$ indicate the usual symmetrization, i.e., the sum over permutations of indices divided by $n!$ ~\cite{1973_MTW}. The first equation describes the dynamics of the background quantities of the fluid, such as the temperature and four-velocity, and includes the backreaction of the nonequilibrium tensor $\zeta_{\mu\nu}$ on the background dynamics. The second and third equations are coupled and describe the evolution of both nonequilibrium tensors.

In this work we only consider tensor perturbations of the fluid and the Minkowski metric, the latter describing the gravitational waves as $g_{\mu\nu} = \eta_{\mu\nu} + h_{\mu\nu}$. Finally, the background evolution can be reduced to the single equation \eqref{First Boltzmann equation}, while the tensor dynamics, after including the metric perturbation, becomes eqs. \eqref{Second and third Boltzmann equation and Einstein equation}.

\section{The $L_{k,l}$ functions}
\label{Section_L functions}

In this appendix we present the general structure of the functions \( L_{k,l}(z) \), defined in Section \ref{Section_background}. These functions appear when factoring out the dependence on $ M$ and $z$ from the coefficients in \eqref{a^k_l functions}
\be
a^k_l=\frac{n_*}{2\pi^2}M^{2+k+l}L_{kl}\left(z\right)
\te
which depend on $z=M/T$ through the functions
\be
\label{L_kl}
L_{k,l}(z)=\int_0^\infty dx\,\sinh^{k+2}(x)\,\cosh^l(x) \, e^{-z\cosh x}.
\te

For $l=0$, these are modified Bessel functions of the second kind
\be
\label{L_kl with l=0}
L_{k,0}(z)=2\sqrt{\frac{2^k}\pi}\Gamma\lc\frac{k+3}{2}\rc z^{-(1+k/2)}K_{1+k/2}(z).
\te
The case $k=0$ is particularly important, as it determines the thermal mass of the fluid
\be
\label{Detailed thermal mass}
M_T^2=\frac{n_*}{2\pi^2}M^2L_{0,0}(z)
\te
and from \eqref{L_kl with l=0} we find
\be
L_{0,0}(z)=\frac{K_1(z)}{z}.
\te

For $l\geq0$ the functions satisfy the identity $L'_{k,l}=-L_{k,l+1}$, which allows us to express them as linear combinations of the Bessel functions by using their recurrence relations. In the asymptotic limit $z\to\infty$ these functions behave as
\be
\label{L_kl asymptotic limit}
L_{k,l}(z)=\sqrt{2^{k+1}}\Gamma\lc\frac{k+3}{2}\rc z^{-\frac{k+3}{2}}e^{-z}
\te
which holds for all $l$, since in this limit the dependence on $l$ becomes of a higher order in $1/z$. For $l<0$ these functions cannot be expressed in terms of Bessel functions and must be computed numerically.

\section{Evaluation of the relaxation time}
\label{Section_Relaxation time calculation}
In this appendix we evaluate the relaxation time in this model. To get an expression for this parameter we evaluate the viscosity of the fluid and its entropy density. We then assume the viscosity-to-entropy ratio $\eta/s$ as a free parameter that verifies the AdS/CFT bound $\eta/s\geq 1/4\pi$. 

The evaluation of the relaxation time involves the assumption that the fluid is close to equilibrium, which implies that $\tau\to 0$. It can be shown that in this limit, the second equation in \eqref{Appendix_Full_Hydro_Eqs} becomes 
\be
\frac13\lc a^2_0-\frac25 a^4_{-2} \rc \sigma^{\mu\nu}+\frac{2}{15}\frac{a^4_{-1}}{\tau}\zeta^{\mu\nu}=0
\te
where $\sigma_{\mu\nu}$ is the shear tensor
\be
\label{Shear tensor}
\sigma^{\mu\nu}=\Delta^\mu_\alpha\Delta^\nu_\beta
\left[ u^{\alpha;\beta}+u^{\beta;\alpha}-\frac23 \Delta^{\alpha\beta}u^\lambda_{;\lambda} \right].
\te
Under this assumptions we evaluate the viscous energy-momentum tensor of the fluid $\Pi^{\mu\nu}=T^{\mu\nu}-T_0^{\mu\nu}$ and match it with the usual expression of the shear viscosity $\Pi_{\mu\nu}=-\eta\sigma^{\mu\nu}$ to get
\be
\eta=\frac{\tau}{3}\lc a^2_0-\frac25 a^4_{-2} \rc.
\te

We now evaluate the equilibrium entropy current \eqref{entropyflux} to get the equilibrium entropy density
\be
s =-u_\mu S^\mu_{0}= a^0_1+\frac{1}{T}a^0_2.
\te

Finally, we get an expression for the relaxation time of the fluid
\be
\tau=3\lc\frac{\eta}{s}\rc \, \frac{a^0_1  + T^{-1} a^0_2}{a^2_0 - \tfrac15 a^4_{-2}}.
\te
By using the properties of the functions $a^k_l$, that are shown in Appendix \ref{Section_L functions}, we get a simpler expression for the relaxation time
\be
\label{Relaxation time-Appendix}
\tau=\lc\frac{\eta}{s}\rc \frac{3}{M}\, \frac{K_3(z)}{L_{2,0} - \tfrac15 L_{4,-2}}
\te
where $K_3(z)$ is a modified Bessel function of the second kind.


\bibliography{references}

@PREAMBLE{
 "\providecommand{\noopsort}[1]{}" 
 # "\providecommand{\singleletter}[1]{#1}%" 
}

@article{huang2015,
  title = {Glasma evolution and {Bose}-{Einstein} condensation with elastic and inelastic collisions},
  author = {Huang, Xu-Guang and Liao, Jinfeng},
  journal = {Phys. Rev. D},
  volume = {91},
  issue = {11},
  pages = {116012},
  numpages = {23},
  year = {2015},
  month = {Jun},
  publisher = {American Physical Society},
  doi = {10.1103/PhysRevD.91.116012},
  url = {https://link.aps.org/doi/10.1103/PhysRevD.91.116012}
}

@article{florkowski2016,
  title = {Separation of elastic and inelastic processes in the relaxation-time approximation for the collision integral},
  author = {Florkowski, Wojciech and Ryblewski, Radoslaw},
  journal = {Phys. Rev. C},
  volume = {93},
  issue = {6},
  pages = {064903},
  numpages = {7},
  year = {2016},
  month = {Jun},
  publisher = {American Physical Society},
  doi = {10.1103/PhysRevC.93.064903},
  url = {https://link.aps.org/doi/10.1103/PhysRevC.93.064903}
}

@article{jeon2025,
  title = {Analytic structure of stress-energy response functions and new {Kubo} formulas},
  author = {Jeon, Sangyong and Czajka, Alina and Hong, Juhee},
  journal = {Phys. Rev. C},
  volume = {112},
  issue = {6},
  pages = {064905},
  numpages = {13},
  year = {2025},
  month = {Dec},
  publisher = {American Physical Society},
  doi = {10.1103/7rrg-2nq3},
  url = {https://link.aps.org/doi/10.1103/7rrg-2nq3}
}

@article{jeon2024,
  title = {Evolution equation for the energy-momentum moments of the nonequilibrium density function and regularized relativistic third-order hydrodynamics},
  author = {Ye, Dasen and Jeon, Sangyong and Gale, Charles},
  journal = {Phys. Rev. C},
  volume = {110},
  issue = {2},
  pages = {024907},
  numpages = {28},
  year = {2024},
  month = {Aug},
  publisher = {American Physical Society},
  doi = {10.1103/PhysRevC.110.024907},
  url = {https://link.aps.org/doi/10.1103/PhysRevC.110.024907}
}

@article{panday2024,
  title = {Causal third-order viscous hydrodynamics within relaxation-time approximation},
  author = {Panday, Pushpa and Jaiswal, Amaresh and Patra, Binoy Krishna},
  journal = {Phys. Rev. D},
  volume = {109},
  issue = {9},
  pages = {096039},
  numpages = {14},
  year = {2024},
  month = {May},
  publisher = {American Physical Society},
  doi = {10.1103/PhysRevD.109.096039},
  url = {https://link.aps.org/doi/10.1103/PhysRevD.109.096039}
}

@article{brito2023,
  title = {Third-order relativistic dissipative fluid dynamics from the method of moments},
  author = {de Brito, Caio V. P. and Denicol, Gabriel S.},
  journal = {Phys. Rev. D},
  volume = {108},
  issue = {9},
  pages = {096020},
  numpages = {15},
  year = {2023},
  month = {Nov},
  publisher = {American Physical Society},
  doi = {10.1103/PhysRevD.108.096020},
  url = {https://link.aps.org/doi/10.1103/PhysRevD.108.096020}
}

@article{brito2022,
  title = {Linear causality and stability of third-order relativistic dissipative fluid dynamics},
  author = {Brito, C. V. and Denicol, G. S.},
  journal = {Phys. Rev. D},
  volume = {105},
  issue = {9},
  pages = {096026},
  numpages = {18},
  year = {2022},
  month = {May},
  publisher = {American Physical Society},
  doi = {10.1103/PhysRevD.105.096026},
  url = {https://link.aps.org/doi/10.1103/PhysRevD.105.096026}
}

@article{jaiswal2013,
  title = {Relativistic third-order dissipative fluid dynamics from kinetic theory},
  author = {Jaiswal, Amaresh},
  journal = {Phys. Rev. C},
  volume = {88},
  issue = {2},
  pages = {021903},
  numpages = {5},
  year = {2013},
  month = {Aug},
  publisher = {American Physical Society},
  doi = {10.1103/PhysRevC.88.021903},
  url = {https://link.aps.org/doi/10.1103/PhysRevC.88.021903}
}

@article{denicol2012,
  title = {Derivation of transient relativistic fluid dynamics from the {Boltzmann} equation},
  author = {Denicol, G. S. and Niemi, H. and Moln\'ar, E. and Rischke, D. H.},
  journal = {Phys. Rev. D},
  volume = {85},
  issue = {11},
  pages = {114047},
  numpages = {22},
  year = {2012},
  month = {Jun},
  publisher = {American Physical Society},
  doi = {10.1103/PhysRevD.85.114047},
  url = {https://link.aps.org/doi/10.1103/PhysRevD.85.114047}
}

@article{struchtrup2003,
    author = {Struchtrup, Henning and Torrilhon, Manuel},
    title = {Regularization of Grad's 13 moment equations: Derivation and linear analysis},
    journal = {Physics of Fluids},
    volume = {15},
    number = {9},
    pages = {2668-2680},
    year = {2003},
    month = {09},
    issn = {1070-6631},
    doi = {10.1063/1.1597472},
    url = {https://doi.org/10.1063/1.1597472},
}

@article{jeon1996,
  title = {From quantum field theory to hydrodynamics: Transport coefficients and effective kinetic theory},
  author = {Jeon, Sangyong and Yaffe, Laurence G.},
  journal = {Phys. Rev. D},
  volume = {53},
  issue = {10},
  pages = {5799--5809},
  numpages = {0},
  year = {1996},
  month = {May},
  publisher = {American Physical Society},
  doi = {10.1103/PhysRevD.53.5799},
  url = {https://link.aps.org/doi/10.1103/PhysRevD.53.5799}
}

@article{jeon1993,
  title = {Computing spectral densities in finite temperature field theory},
  author = {Jeon, Sangyong},
  journal = {Phys. Rev. D},
  volume = {47},
  issue = {10},
  pages = {4586--4607},
  numpages = {0},
  year = {1993},
  month = {May},
  publisher = {American Physical Society},
  doi = {10.1103/PhysRevD.47.4586},
  url = {https://link.aps.org/doi/10.1103/PhysRevD.47.4586}
}

@article{calzetta1988,
  title = {Nonequilibrium quantum fields: Closed-time-path effective action, Wigner function, and Boltzmann equation},
  author = {Calzetta, E. and Hu, B. L.},
  journal = {Phys. Rev. D},
  volume = {37},
  issue = {10},
  pages = {2878--2900},
  numpages = {0},
  year = {1988},
  month = {May},
  publisher = {American Physical Society},
  doi = {10.1103/PhysRevD.37.2878},
  url = {https://link.aps.org/doi/10.1103/PhysRevD.37.2878}
}

@article{calzetta2000,
  title = {Hydrodynamic transport functions from quantum kinetic field theory},
  author = {Calzetta, E. A. and Hu, B. L. and Ramsey, S. A.},
  journal = {Phys. Rev. D},
  volume = {61},
  issue = {12},
  pages = {125013},
  numpages = {20},
  year = {2000},
  month = {May},
  publisher = {American Physical Society},
  doi = {10.1103/PhysRevD.61.125013},
  url = {https://link.aps.org/doi/10.1103/PhysRevD.61.125013}
}

@article{jeon1995,
  title = {Hydrodynamic transport coefficients in relativistic scalar field theory},
  author = {Jeon, Sangyong},
  journal = {Phys. Rev. D},
  volume = {52},
  issue = {6},
  pages = {3591--3642},
  numpages = {0},
  year = {1995},
  month = {Sep},
  publisher = {American Physical Society},
  doi = {10.1103/PhysRevD.52.3591},
  url = {https://link.aps.org/doi/10.1103/PhysRevD.52.3591}
}

@book{Schutz_2022, place={Cambridge}, edition={3}, title={A First Course in General Relativity}, publisher={Cambridge University Press}, author={Schutz, Bernard}, year={2022}}

@article{Calzetta_1998,
doi = {10.1088/0264-9381/15/3/015},
url = {https://doi.org/10.1088/0264-9381/15/3/015},
year = {1998},
month = {mar},
publisher = {},
volume = {15},
number = {3},
pages = {653},
author = {Esteban Calzetta},
title = {Relativistic fluctuating hydrodynamics},
journal = {Classical and Quantum Gravity}
}

@article{Garcia_mambrini_2023,
doi = {10.1088/1475-7516/2023/12/028},
url = {https://doi.org/10.1088/1475-7516/2023/12/028},
year = {2023},
month = {dec},
publisher = {IOP Publishing},
volume = {2023},
number = {12},
pages = {028},
author = {Garcia, Marcos A.G. and Gross, Mathieu and Mambrini, Yann and Olive, Keith A. and Pierre, Mathias and Yoon, Jong-Hyun},
title = {Effects of fragmentation on post-inflationary reheating},
journal = {Journal of Cosmology and Astroparticle Physics}
}

@book{chapman_cowling,
  title     = {The Mathematical Theory of Non-uniform Gases: An Account of the Kinetic Theory of Viscosity, Thermal Conduction and Diffusion in Gases},
  author    = {Chapman, Sydney and Cowling, T. G.},
  edition   = {3rd},
  series    = {Cambridge Mathematical Library},
  publisher = {Cambridge University Press},
  year      = {1991},
  month     = feb,
  isbn      = {9780521408448}
}

@article{bhatnagar_1954,
  title = {A Model for Collision Processes in Gases. I. Small Amplitude Processes in Charged and Neutral One-Component Systems},
  author = {Bhatnagar, P. L. and Gross, E. P. and Krook, M.},
  journal = {Phys. Rev.},
  volume = {94},
  issue = {3},
  pages = {511--525},
  numpages = {0},
  year = {1954},
  month = {May},
  publisher = {American Physical Society},
  doi = {10.1103/PhysRev.94.511},
  url = {https://link.aps.org/doi/10.1103/PhysRev.94.511}
}

@article{geroch_1990,
  title = {Dissipative relativistic fluid theories of divergence type},
  author = {Geroch, Robert and Lindblom, Lee},
  journal = {Phys. Rev. D},
  volume = {41},
  issue = {6},
  pages = {1855--1861},
  numpages = {0},
  year = {1990},
  month = {Mar},
  publisher = {American Physical Society},
  doi = {10.1103/PhysRevD.41.1855},
  url = {https://link.aps.org/doi/10.1103/PhysRevD.41.1855}
}

@Article{Miron-Granese_jhep_2020,
author={Miron-Granese, Nahuel
and Kandus, Alejandra
and Calzetta, Esteban},
title={Nonlinear fluctuations in relativistic causal fluids},
journal={Journal of High Energy Physics},
year={2020},
month={Jul},
day={10},
volume={2020},
number={7},
pages={64},
issn={1029-8479},
doi={10.1007/JHEP07(2020)064},
url={https://doi.org/10.1007/JHEP07(2020)064}
}

@article{MITRA2025100054,
title = {Causality and stability in relativistic hydrodynamics},
journal = {Journal of Subatomic Particles and Cosmology},
volume = {3},
pages = {100054},
year = {2025},
issn = {3050-4805},
doi = {https://doi.org/10.1016/j.jspc.2025.100054},
url = {https://www.sciencedirect.com/science/article/pii/S3050480525000342},
author = {Sukanya Mitra}
}

@article{Boyanovsky_1996,
   title={Analytic and numerical study of preheating dynamics},
   volume={54},
   ISSN={1089-4918},
   url={http://dx.doi.org/10.1103/PhysRevD.54.7570},
   DOI={10.1103/physrevd.54.7570},
   number={12},
   journal={Physical Review D},
   publisher={American Physical Society (APS)},
   author={Boyanovsky, D. and de Vega, H. J. and Holman, R. and Salgado, J. F. J.},
   year={1996},
   month=dec, pages={7570–7598} }

@book{berselli2006mathematics,
  title={Mathematics of large eddy simulation of turbulent flows},
  author={Berselli, Luigi C and Iliescu, Traian and Layton, William J},
  year={2006},
  publisher={Springer}
}

@book{sagaut2006large,
  title={Large eddy simulation for incompressible flows: an introduction},
  author={Sagaut, Pierre},
  year={2006},
  publisher={Springer}
}

@article{Florkowski_2018,
doi = {10.1088/1361-6633/aaa091},
url = {https://dx.doi.org/10.1088/1361-6633/aaa091},
year = {2018},
month = {feb},
publisher = {IOP Publishing},
volume = {81},
number = {4},
pages = {046001},
author = {Florkowski, Wojciech and Heller, Michal P and Spaliński, Michał},
title = {New theories of relativistic hydrodynamics in the LHC era},
journal = {Reports on Progress in Physics},
}

@article{STRICKLAND_2015,
   title={Thermalization and isotropization in heavy-ion collisions},
   volume={84},
   ISSN={0973-7111},
   url={http://dx.doi.org/10.1007/s12043-015-0972-1},
   DOI={10.1007/s12043-015-0972-1},
   number={5},
   journal={Pramana},
   publisher={Springer Science and Business Media LLC},
   author={Strickland, Michael},
   year={2015},
   month=may, pages={671–684} }

@book{Romatschke_Romatschke_2019, place={Cambridge}, series={Cambridge Monographs on Mathematical Physics}, title={Relativistic Fluid Dynamics In and Out of Equilibrium: And Applications to Relativistic Nuclear Collisions}, publisher={Cambridge University Press}, author={Romatschke, Paul and Romatschke, Ulrike}, year={2019}, collection={Cambridge Monographs on Mathematical Physics}}

@article{Heller_2015,
  title = {Hydrodynamics Beyond the Gradient Expansion: Resurgence and Resummation},
  author = {Heller, Michal P. and Spali\ifmmode \acute{n}\else \'{n}\fi{}ski, Micha\l{}},
  journal = {Phys. Rev. Lett.},
  volume = {115},
  issue = {7},
  pages = {072501},
  numpages = {5},
  year = {2015},
  month = {Aug},
  publisher = {American Physical Society},
  doi = {10.1103/PhysRevLett.115.072501},
  url = {https://link.aps.org/doi/10.1103/PhysRevLett.115.072501}
}

@article{heller_2022,
  title = {Relativistic Hydrodynamics: A Singulant Perspective},
  author = {Heller, Michal P. and Serantes, Alexandre and Spali\ifmmode \acute{n}\else \'{n}\fi{}ski, Micha\l{} and Svensson, Viktor and Withers, Benjamin},
  journal = {Phys. Rev. X},
  volume = {12},
  issue = {4},
  pages = {041010},
  numpages = {25},
  year = {2022},
  month = {Oct},
  publisher = {American Physical Society},
  doi = {10.1103/PhysRevX.12.041010},
  url = {https://link.aps.org/doi/10.1103/PhysRevX.12.041010}
}

@Article{Soloviev_2022,
author={Soloviev, Alexander},
title={Hydrodynamic attractors in heavy ion collisions: a review},
journal={The European Physical Journal C},
year={2022},
month={Apr},
day={12},
volume={82},
number={4},
pages={319},
issn={1434-6052},
doi={10.1140/epjc/s10052-022-10282-4},
url={https://doi.org/10.1140/epjc/s10052-022-10282-4}
}

@article{nahuel_claudia_2025,
  title = {Bias in the tensor-to-scalar ratio from self-interacting dark radiation},
  author = {Mir\'on-Granese, Nahuel and Sc\'occola, Claudia G.},
  journal = {Phys. Rev. D},
  volume = {112},
  issue = {12},
  pages = {123516},
  numpages = {10},
  year = {2025},
  month = {Dec},
  publisher = {American Physical Society},
  doi = {10.1103/z87v-4dp4},
  url = {https://link.aps.org/doi/10.1103/z87v-4dp4}
}

@article{nahuel_esteban_2018,
  title = {Primordial gravitational waves amplification from causal fluids},
  author = {Miron-Granese, Nahuel and Calzetta, Esteban},
  journal = {Phys. Rev. D},
  volume = {97},
  issue = {2},
  pages = {023517},
  numpages = {14},
  year = {2018},
  month = {Jan},
  publisher = {American Physical Society},
  doi = {10.1103/PhysRevD.97.023517},
  url = {https://link.aps.org/doi/10.1103/PhysRevD.97.023517}
}

@article{2021_Miron-Granese,
   title={Relativistic viscous effects on the primordial gravitational waves spectrum},
   volume={2021},
   ISSN={1475-7516},
   url={http://dx.doi.org/10.1088/1475-7516/2021/06/008},
   DOI={10.1088/1475-7516/2021/06/008},
   number={06},
   journal={Journal of Cosmology and Astroparticle Physics},
   publisher={IOP Publishing},
   author={Mirón-Granese, Nahuel},
   year={2021},
   month=jun, pages={008} }

@article{baym_2017,
  title = {Damping of gravitational waves by matter},
  author = {Baym, Gordon and Patil, Subodh P. and Pethick, C. J.},
  journal = {Phys. Rev. D},
  volume = {96},
  issue = {8},
  pages = {084033},
  numpages = {10},
  year = {2017},
  month = {Oct},
  publisher = {American Physical Society},
  doi = {10.1103/PhysRevD.96.084033},
  url = {https://link.aps.org/doi/10.1103/PhysRevD.96.084033}
}

@book{1976_Landau,
   author       = "L. Landau and L. Lifshits",
   year= "1976",
   title="Mechanics",
   publisher="Pergamon Press"
}

@book{2008_Calzetta_Hu,
  author    = {Calzetta, Esteban A. and Hu, Bei-Lok B.},
  title     = {Nonequilibrium Quantum Field Theory},
  publisher = {Cambridge University Press},
  series    = {Cambridge Monographs on Mathematical Physics},
  year      = {2008},
  isbn      = {978-0-521-64168-5},
}

@article{1997_KLS,
   title={Towards the theory of reheating after inflation},
   volume={56},
   ISSN={1089-4918},
   url={http://dx.doi.org/10.1103/PhysRevD.56.3258},
   DOI={10.1103/physrevd.56.3258},
   number={6},
   journal={Physical Review D},
   publisher={American Physical Society (APS)},
   author={Kofman, Lev and Linde, Andrei and Starobinsky, Alexei A.},
   year={1997},
   month=sep, pages={3258–3295} }

@article{2020_Cantarutti_Calzetta,
author = {Cantarutti, Lucas and Calzetta, Esteban},
title = {Dissipative-type theories for {Bjorken} and {Gubser} flows},
journal = {International Journal of Modern Physics A},
volume = {35},
number = {14},
pages = {2050074},
year = {2020},
doi = {10.1142/S0217751X20500748},
URL = {https://doi.org/10.1142/S0217751X20500748},
}

@Article{2021_Hindmarsh,
	title={{Phase transitions in the early universe}},
	author={Mark Hindmarsh and Marvin Lüben and Johannes Lumma and Martin Pauly},
	journal={SciPost Phys. Lect. Notes},
	pages={24},
	year={2021},
	publisher={SciPost},
	doi={10.21468/SciPostPhysLectNotes.24},
	url={https://scipost.org/10.21468/SciPostPhysLectNotes.24},
}

@article{2021_Perna_Calzetta,
  title = {Linearized dispersion relations in viscous relativistic hydrodynamics},
  author = {Perna, Guillermo and Calzetta, Esteban},
  journal = {Phys. Rev. D},
  volume = {104},
  issue = {9},
  pages = {096005},
  numpages = {20},
  year = {2021},
  publisher = {American Physical Society},
  doi = {10.1103/PhysRevD.104.096005},
  url = {https://link.aps.org/doi/10.1103/PhysRevD.104.096005}
}

@article{2020_McDonough,
title = {The cosmological heavy ion collider: Fast thermalization after cosmic inflation},
journal = {Physics Letters B},
volume = {809},
pages = {135755},
year = {2020},
issn = {0370-2693},
doi = {https://doi.org/10.1016/j.physletb.2020.135755},
url = {https://www.sciencedirect.com/science/article/pii/S037026932030558X},
author = {Evan McDonough}
}

@article{2001_Policastro,
  title = {Shear Viscosity of Strongly Coupled $N\phantom{\rule{0ex}{0ex}}=\phantom{\rule{0ex}{0ex}}4$ Supersymmetric Yang-Mills Plasma},
  author = {Policastro, G. and Son, D. T. and Starinets, A. O.},
  journal = {Phys. Rev. Lett.},
  volume = {87},
  issue = {8},
  pages = {081601},
  numpages = {4},
  year = {2001},
  month = {Aug},
  publisher = {American Physical Society},
  doi = {10.1103/PhysRevLett.87.081601},
  url = {https://link.aps.org/doi/10.1103/PhysRevLett.87.081601}
}

@article{2001_Cremonini,
author = {Cremonini, Sera},
title = {THE SHEAR VISCOSITY TO ENTROPY RATIO: A STATUS REPORT},
journal = {Modern Physics Letters B},
volume = {25},
number = {23},
pages = {1867-1888},
year = {2011},
doi = {10.1142/S0217984911027315},

URL = { 
    
        https://doi.org/10.1142/S0217984911027315
    
    

},

}

@article{2010_Martin,
  title = {First CMB constraints on the inflationary reheating temperature},
  author = {Martin, J\'er\^ome and Ringeval, Christophe},
  journal = {Phys. Rev. D},
  volume = {82},
  issue = {2},
  pages = {023511},
  numpages = {17},
  year = {2010},
  month = {Jul},
  publisher = {American Physical Society},
  doi = {10.1103/PhysRevD.82.023511},
  url = {https://link.aps.org/doi/10.1103/PhysRevD.82.023511}
}

@article{2015_Martin,
  title = {Observing Inflationary Reheating},
  author = {Martin, J\'er\^ome and Ringeval, Christophe and Vennin, Vincent},
  journal = {Phys. Rev. Lett.},
  volume = {114},
  issue = {8},
  pages = {081303},
  numpages = {5},
  year = {2015},
  month = {Feb},
  publisher = {American Physical Society},
  doi = {10.1103/PhysRevLett.114.081303},
  url = {https://link.aps.org/doi/10.1103/PhysRevLett.114.081303}
}

@article{2018_Caprini,
   title={Cosmological backgrounds of gravitational waves},
   volume={35},
   ISSN={1361-6382},
   url={http://dx.doi.org/10.1088/1361-6382/aac608},
   DOI={10.1088/1361-6382/aac608},
   number={16},
   journal={Classical and Quantum Gravity},
   publisher={IOP Publishing},
   author={Caprini, Chiara and Figueroa, Daniel G},
   year={2018},
   month=jul, pages={163001} 
}

@misc{2019_Lozanov,
      title={Lectures on Reheating after Inflation}, 
      author={Kaloian D. Lozanov},
      year={2019},
      eprint={1907.04402},
      archivePrefix={arXiv},
      primaryClass={astro-ph.CO},
      url={https://arxiv.org/abs/1907.04402}, 
}

@article{2014_Amin,
author = {Amin, Mustafa A. and Hertzberg, Mark P. and Kaiser, David I. and Karouby, Johanna},
title = {Nonperturbative dynamics of reheating after inflation: A review},
journal = {International Journal of Modern Physics D},
volume = {24},
number = {01},
pages = {1530003},
year = {2015},
doi = {10.1142/S0218271815300037},

URL = { 
    
        https://doi.org/10.1142/S0218271815300037
    
    

}
}

@misc{2024_Mishra,
      title={Cosmic Inflation: Background dynamics, Quantum fluctuations and Reheating}, 
      author={Swagat S. Mishra},
      year={2024},
      eprint={2403.10606},
      archivePrefix={arXiv},
      primaryClass={gr-qc},
      url={https://arxiv.org/abs/2403.10606}, 
}

@article{1997_Khlebnikov,
   title={Relic gravitational waves produced after preheating},
   volume={56},
   ISSN={1089-4918},
   url={http://dx.doi.org/10.1103/PhysRevD.56.653},
   DOI={10.1103/physrevd.56.653},
   number={2},
   journal={Physical Review D},
   publisher={American Physical Society (APS)},
   author={Khlebnikov, S. and Tkachev, I.},
   year={1997},
   month=jul, pages={653–660} }

@article{2006_Easther,
   title={Stochastic gravitational wave production after inflation},
   volume={2006},
   ISSN={1475-7516},
   url={http://dx.doi.org/10.1088/1475-7516/2006/04/010},
   DOI={10.1088/1475-7516/2006/04/010},
   number={04},
   journal={Journal of Cosmology and Astroparticle Physics},
   publisher={IOP Publishing},
   author={Easther, Richard and Lim, Eugene A},
   year={2006},
   month=apr, pages={010–010} }

@article{2008_Bellido,
  title = {Gravitational wave background from reheating after hybrid inflation},
  author = {Garc\'{\i}a-Bellido, Juan and Figueroa, Daniel G. and Sastre, Alfonso},
  journal = {Phys. Rev. D},
  volume = {77},
  issue = {4},
  pages = {043517},
  numpages = {23},
  year = {2008},
  month = {Feb},
  publisher = {American Physical Society},
  doi = {10.1103/PhysRevD.77.043517},
  url = {https://link.aps.org/doi/10.1103/PhysRevD.77.043517}
}

@article{1994_KLS,
  author = {L. Kofman and A. Linde and A. Starobinsky},
  title = {Reheating after Inflation},
  journal = {Phys. Rev. Lett.},
  volume = {73},
  pages = {3195--3198},
  year = {1994},
  doi = {10.1103/PhysRevLett.73.3195}
}

@article{1990_Traschen,
  author = {J. Traschen and R. Brandenberger},
  title = {Particle production during out-of-equilibrium phase transitions},
  journal = {Phys. Rev. D},
  volume = {42},
  pages = {2491--2504},
  year = {1990},
  doi = {10.1103/PhysRevD.42.2491}
}

@ARTICLE{1990_Dolgov,
       author = {{Dolgov}, A.~D. and {Kirilova}, D.~P.},
        title = "{Production of particles by a variable scalar field}",
      journal = {Soviet Journal of Nuclear Physics},
         year = 1990,
        month = jan,
       volume = {51},
        pages = {172-177},
       adsurl = {https://ui.adsabs.harvard.edu/abs/1990SvJNP..51..172D}
}

@article{1995_Shtanov,
   title={Universe reheating after inflation},
   volume={51},
   ISSN={0556-2821},
   url={http://dx.doi.org/10.1103/PhysRevD.51.5438},
   DOI={10.1103/physrevd.51.5438},
   number={10},
   journal={Physical Review D},
   publisher={American Physical Society (APS)},
   author={Shtanov, Y. and Traschen, J. and Brandenberger, R.},
   year={1995},
   month=may, pages={5438–5455} }

@article{1997_Ramsey,
  author = {S. A. Ramsey and B. L. Hu},
  title = {Nonequilibrium inflaton dynamics and reheating: Back reaction of parametric particle creation and curved spacetime effects},
  journal = {Physical Review D},
  volume = {56},
  number = {2},
  pages = {678--707},
  year = {1997},
  doi = {10.1103/PhysRevD.56.678}
}

@article{1996_Khlebnikov,
   title={Classical Decay of the Inflaton},
   volume={77},
   ISSN={1079-7114},
   url={http://dx.doi.org/10.1103/PhysRevLett.77.219},
   DOI={10.1103/physrevlett.77.219},
   number={2},
   journal={Physical Review Letters},
   publisher={American Physical Society (APS)},
   author={Khlebnikov, S. Yu. and Tkachev, I. I.},
   year={1996},
   month=jul, pages={219–222} }

@article{1997_Khlebnikov_Tkachev,
  title = {Resonant Decay of Cosmological Bose Condensates},
  author = {Khlebnikov, S. and Tkachev, I.},
  journal = {Phys. Rev. Lett.},
  volume = {79},
  issue = {9},
  pages = {1607--1610},
  numpages = {0},
  year = {1997},
  month = {Sep},
  publisher = {American Physical Society},
  doi = {10.1103/PhysRevLett.79.1607},
  url = {https://link.aps.org/doi/10.1103/PhysRevLett.79.1607}
}

@article{2006_Podolsky,
  title = {Equation of state and beginning of thermalization after preheating},
  author = {Podolsky, Dmitry and Felder, Gary N. and Kofman, Lev and Peloso, Marco},
  journal = {Phys. Rev. D},
  volume = {73},
  issue = {2},
  pages = {023501},
  numpages = {15},
  year = {2006},
  month = {Jan},
  publisher = {American Physical Society},
  doi = {10.1103/PhysRevD.73.023501},
  url = {https://link.aps.org/doi/10.1103/PhysRevD.73.023501}
}

@article{2023_Garcia,
  author = {M. A. Garcia and M. Pierre},
  title = {Reheating after inflaton fragmentation},
  journal = {Journal of Cosmology and Astroparticle Physics},
  volume = {2023},
  number = {11},
  pages = {004},
  year = {2023},
  doi = {10.1088/1475-7516/2023/11/004}
}

@article{2000_Felder,
title = {LATTICEEASY: A program for lattice simulations of scalar fields in an expanding universe},
journal = {Computer Physics Communications},
volume = {178},
number = {12},
pages = {929-932},
year = {2008},
issn = {0010-4655},
doi = {https://doi.org/10.1016/j.cpc.2008.02.009},
url = {https://www.sciencedirect.com/science/article/pii/S001046550800091X},
author = {Gary Felder and Igor Tkachev}
}

@article{2008_Frolov,
   title={DEFROST: a new code for simulating preheating after inflation},
   volume={2008},
   ISSN={1475-7516},
   url={http://dx.doi.org/10.1088/1475-7516/2008/11/009},
   DOI={10.1088/1475-7516/2008/11/009},
   number={11},
   journal={Journal of Cosmology and Astroparticle Physics},
   publisher={IOP Publishing},
   author={Frolov, Andrei V},
   year={2008},
   month=nov, pages={009} }

@article{2024_Figueroa,
   title={Present and future of CosmoLattice},
   volume={87},
   ISSN={1361-6633},
   url={http://dx.doi.org/10.1088/1361-6633/ad616a},
   DOI={10.1088/1361-6633/ad616a},
   number={9},
   journal={Reports on Progress in Physics},
   publisher={IOP Publishing},
   author={Figueroa, Daniel G and Florio, Adrien and Torrenti, Francisco},
   year={2024},
   month=aug, pages={094901} }

@article{2017_Figueroa,
   title={Gravitational wave production from preheating: parameter dependence},
   volume={2017},
   ISSN={1475-7516},
   url={http://dx.doi.org/10.1088/1475-7516/2017/10/057},
   DOI={10.1088/1475-7516/2017/10/057},
   number={10},
   journal={Journal of Cosmology and Astroparticle Physics},
   publisher={IOP Publishing},
   author={Figueroa, Daniel G. and Torrentí, Francisco},
   year={2017},
   month=oct, pages={057–057} }

@article{2002_Grana,
  title = {Reheating and turbulence},
  author = {Gra\~na, Mariana and Calzetta, Esteban},
  journal = {Phys. Rev. D},
  volume = {65},
  issue = {6},
  pages = {063522},
  numpages = {9},
  year = {2002},
  month = {Mar},
  publisher = {American Physical Society},
  doi = {10.1103/PhysRevD.65.063522},
  url = {https://link.aps.org/doi/10.1103/PhysRevD.65.063522}
}

@article{2003_Micha,
  title = {Relativistic Turbulence: A Long Way from Preheating to Equilibrium},
  author = {Micha, Raphael and Tkachev, Igor I.},
  journal = {Phys. Rev. Lett.},
  volume = {90},
  issue = {12},
  pages = {121301},
  numpages = {4},
  year = {2003},
  month = {Mar},
  publisher = {American Physical Society},
  doi = {10.1103/PhysRevLett.90.121301},
  url = {https://link.aps.org/doi/10.1103/PhysRevLett.90.121301}
}

@article{2004_Micha,
  title = {Turbulent thermalization},
  author = {Micha, Raphael and Tkachev, Igor I.},
  journal = {Phys. Rev. D},
  volume = {70},
  issue = {4},
  pages = {043538},
  numpages = {25},
  year = {2004},
  month = {Aug},
  publisher = {American Physical Society},
  doi = {10.1103/PhysRevD.70.043538},
  url = {https://link.aps.org/doi/10.1103/PhysRevD.70.043538}
}

@article{2004_Berges,
  title = {Prethermalization},
  author = {Berges, J. and Bors\'anyi, Sz. and Wetterich, C.},
  journal = {Phys. Rev. Lett.},
  volume = {93},
  issue = {14},
  pages = {142002},
  numpages = {4},
  year = {2004},
  month = {Sep},
  publisher = {American Physical Society},
  doi = {10.1103/PhysRevLett.93.142002},
  url = {https://link.aps.org/doi/10.1103/PhysRevLett.93.142002}
}

@article{2015_Cook,
   title={Reheating predictions in single field inflation},
   volume={2015},
   ISSN={1475-7516},
   url={http://dx.doi.org/10.1088/1475-7516/2015/04/047},
   DOI={10.1088/1475-7516/2015/04/047},
   number={04},
   journal={Journal of Cosmology and Astroparticle Physics},
   publisher={IOP Publishing},
   author={Cook, Jessica L. and Dimastrogiovanni, Emanuela and Easson, Damien A. and Krauss, Lawrence M.},
   year={2015},
   month=apr, pages={047–047} }

@article{2024_Martin,
   title={Cosmic Inflation at the crossroads},
   volume={2024},
   ISSN={1475-7516},
   url={http://dx.doi.org/10.1088/1475-7516/2024/07/087},
   DOI={10.1088/1475-7516/2024/07/087},
   number={07},
   journal={Journal of Cosmology and Astroparticle Physics},
   publisher={IOP Publishing},
   author={Martin, Jérôme and Ringeval, Christophe and Vennin, Vincent},
   year={2024},
   month=jul, pages={087} }

@article{2007_Easther,
  title = {Gravitational Wave Production at the End of Inflation},
  author = {Easther, Richard and Giblin, John T. and Lim, Eugene A.},
  journal = {Phys. Rev. Lett.},
  volume = {99},
  issue = {22},
  pages = {221301},
  numpages = {4},
  year = {2007},
  month = {Nov},
  publisher = {American Physical Society},
  doi = {10.1103/PhysRevLett.99.221301},
  url = {https://link.aps.org/doi/10.1103/PhysRevLett.99.221301}
}

@article{2007_Dufaux,
  title = {Theory and numerics of gravitational waves from preheating after inflation},
  author = {Dufaux, Jean-Francois and Bergman, Amanda and Felder, Gary and Kofman, Lev and Uzan, Jean-Philippe},
  journal = {Phys. Rev. D},
  volume = {76},
  issue = {12},
  pages = {123517},
  numpages = {24},
  year = {2007},
  month = {Dec},
  publisher = {American Physical Society},
  doi = {10.1103/PhysRevD.76.123517},
  url = {https://link.aps.org/doi/10.1103/PhysRevD.76.123517}
}

@article{2010_Allahverdi,
   title={Reheating in Inflationary Cosmology: Theory and Applications},
   volume={60},
   ISSN={1545-4134},
   url={http://dx.doi.org/10.1146/annurev.nucl.012809.104511},
   DOI={10.1146/annurev.nucl.012809.104511},
   number={1},
   journal={Annual Review of Nuclear and Particle Science},
   publisher={Annual Reviews},
   author={Allahverdi, Rouzbeh and Brandenberger, Robert and Cyr-Racine, Francis-Yan and Mazumdar, Anupam},
   year={2010},
   month=nov, pages={27–51} }

@article{2006_Bassett,
   title={Inflation dynamics and reheating},
   volume={78},
   ISSN={1539-0756},
   url={http://dx.doi.org/10.1103/RevModPhys.78.537},
   DOI={10.1103/revmodphys.78.537},
   number={2},
   journal={Reviews of Modern Physics},
   publisher={American Physical Society (APS)},
   author={Bassett, Bruce A. and Tsujikawa, Shinji and Wands, David},
   year={2006},
   month=may, pages={537–589} }

@article{2024_Kolb,
  author       = {Kolb, Edward W. and Long, Andrew J.},
  title        = {Cosmological gravitational particle production and its implications for cosmological relics},
  journal      = {Reviews of Modern Physics},
  year         = {2024},
  volume       = {96},
  number       = {4},
  pages        = {045005},
  doi          = {10.1103/RevModPhys.96.045005}
}

@article{2025_Roshan,
  author       = {Roshan, R. and White, G.},
  title        = {Using gravitational waves to see the first second of the Universe},
  journal      = {Reviews of Modern Physics},
  year         = {2025},
  volume       = {97},
  number       = {1},
  pages        = {015001},
  doi          = {10.1103/RevModPhys.97.015001}
}

@article{2024_deBrito,
  author       = {de Brito, C. V. and Denicol, G. S.},
  title        = {Method of moments for a relativistic single-component gas},
  journal      = {Physical Review D},
  year         = {2024},
  volume       = {110},
  number       = {3},
  pages        = {036017},
  doi          = {10.1103/PhysRevD.110.036017}
}

@article{2020_Figueroa,
doi = {10.1088/1475-7516/2021/04/035},
url = {https://doi.org/10.1088/1475-7516/2021/04/035},
year = {2021},
month = {apr},
publisher = {IOP Publishing},
volume = {2021},
number = {04},
pages = {035},
author = {Figueroa, Daniel G. and Florio, Adrien and Torrenti, Francisco and Valkenburg, Wessel},
title = {The art of simulating the early universe.
Part I. Integration techniques and canonical cases},
journal = {Journal of Cosmology and Astroparticle Physics}
}

@article{2023_Figueroa,
   title={CosmoLattice: A modern code for lattice simulations of scalar and gauge field dynamics in an expanding universe},
   volume={283},
   ISSN={0010-4655},
   url={http://dx.doi.org/10.1016/j.cpc.2022.108586},
   DOI={10.1016/j.cpc.2022.108586},
   journal={Computer Physics Communications},
   publisher={Elsevier BV},
   author={Figueroa, Daniel G. and Florio, Adrien and Torrenti, Francisco and Valkenburg, Wessel},
   year={2023},
   month=feb, pages={108586} }

@book{2012_Teschl,
  author    = {Gerald Teschl},
  title     = {Ordinary Differential Equations and Dynamical Systems},
  series    = {Graduate Studies in Mathematics},
  volume    = {140},
  publisher = {American Mathematical Society},
  year      = {2012},
  isbn      = {978-0-8218-8328-0},
  pages     = {356}
}

@article{2022_Calzetta,
  author       = {Calzetta, Esteban},
  title        = {Steady asymptotic equilibria in conformal relativistic fluids},
  journal      = {Physical Review D},
  year         = {2022},
  volume       = {105},
  number       = {3},
  pages        = {036013},
  doi          = {10.1103/PhysRevD.105.036013}
}

@article{1974_Anderson,
  author    = {Anderson, J. L. and Witting, H. R.},
  title     = {A Relativistic Relaxation-Time Model for the Boltzmann Equation},
  journal   = {Physica},
  year      = {1974},
  volume    = {74},
  pages     = {466--488},
  doi       = {10.1016/0031-8914(74)90355-3}
}

@article{1974b_Anderson,
  author    = {Anderson, J. L. and Witting, H. R.},
  title     = {Relativistic Quantum Transport Coefficients},
  journal   = {Physica},
  year      = {1974},
  volume    = {74},
  pages     = {489--495},
  doi       = {10.1016/0031-8914(74)90356-5}
}

@article{1997_Greene,
   title={Structure of resonance in preheating after inflation},
   volume={56},
   ISSN={1089-4918},
   url={http://dx.doi.org/10.1103/PhysRevD.56.6175},
   DOI={10.1103/physrevd.56.6175},
   number={10},
   journal={Physical Review D},
   publisher={American Physical Society (APS)},
   author={Greene, Patrick B. and Kofman, Lev and Linde, Andrei and Starobinsky, Alexei A.},
   year={1997},
   month=nov, pages={6175–6192} }

@article{2018_Lozanov,
  title = {Self-resonance after inflation: Oscillons, transients, and radiation domination},
  author = {Lozanov, Kaloian D. and Amin, Mustafa A.},
  journal = {Phys. Rev. D},
  volume = {97},
  issue = {2},
  pages = {023533},
  numpages = {22},
  year = {2018},
  month = {Jan},
  publisher = {American Physical Society},
  doi = {10.1103/PhysRevD.97.023533},
  url = {https://link.aps.org/doi/10.1103/PhysRevD.97.023533}
}

@book{1973_MTW,
  author    = {Misner, Charles W. and Thorne, Kip S. and Wheeler, John A.},
  title     = {Gravitation},
  publisher = {W. H. Freeman},
  address   = {San Francisco},
  year      = {1973},
  isbn      = {0-7167-0344-0},
  pages     = {1279}
}

@article{Calzetta_2021_prd,
  title = {Fully developed relativistic turbulence},
  author = {Calzetta, Esteban},
  journal = {Phys. Rev. D},
  volume = {103},
  issue = {5},
  pages = {056018},
  numpages = {13},
  year = {2021},
  month = {Mar},
  publisher = {American Physical Society},
  doi = {10.1103/PhysRevD.103.056018},
  url = {https://link.aps.org/doi/10.1103/PhysRevD.103.056018}
}

@book{Lifshitz1980StatPhys2,
  title        = {Statistical Physics, Part 2},
  author       = {Lifshitz, E. M. and Pitaevskii, L. P.},
  series       = {Course of Theoretical Physics},
  volume       = {9},
  year         = {1980},
  publisher    = {Pergamon Press},
  address      = {Oxford},
  note         = {Translated by J. B. Sykes and M. J. Kearsley},
  isbn         = {978-0080230379}
}

@article{Touchette_2009,
   title={The large deviation approach to statistical mechanics},
   volume={478},
   ISSN={0370-1573},
   url={http://dx.doi.org/10.1016/j.physrep.2009.05.002},
   DOI={10.1016/j.physrep.2009.05.002},
   number={1–3},
   journal={Physics Reports},
   publisher={Elsevier BV},
   author={Touchette, Hugo},
   year={2009},
   month=jul, pages={1–69} }

@book{2008_Maggiore,
  author    = {Michele Maggiore},
  title     = {Gravitational Waves: Volume 1: Theory and Experiments},
  publisher = {Oxford University Press},
  year      = {2008},
  isbn      = {9780198570745}
}

@book{2018_Maggiore,
  author    = {Michele Maggiore},
  title     = {Gravitational Waves: Volume 2: Astrophysics and Cosmology},
  publisher = {Oxford University Press},
  year      = {2018},
  isbn      = {9780198570899}
}

@article{2022_Miron-Granese,
   title={Field Theory Approaches to Relativistic Hydrodynamics},
   volume={24},
   ISSN={1099-4300},
   url={http://dx.doi.org/10.3390/e24121790},
   DOI={10.3390/e24121790},
   number={12},
   journal={Entropy},
   publisher={MDPI AG},
   author={Mirón Granese, Nahuel and Kandus, Alejandra and Calzetta, Esteban},
   year={2022},
   month=dec, pages={1790} }

@article{1989_Calzetta,
title = {Spinodal decomposition in quantum field theory},
journal = {Annals of Physics},
volume = {190},
number = {1},
pages = {32-58},
year = {1989},
issn = {0003-4916},
doi = {https://doi.org/10.1016/0003-4916(89)90260-1},
url = {https://www.sciencedirect.com/science/article/pii/0003491689902601},
author = {Esteban Calzetta}
}

@article{2001_Felder,
  title = {Dynamics of Symmetry Breaking and Tachyonic Preheating},
  author = {Felder, Gary and Garc\'{\i}a-Bellido, Juan and Greene, Patrick B. and Kofman, Lev and Linde, Andrei and Tkachev, Igor},
  journal = {Phys. Rev. Lett.},
  volume = {87},
  issue = {1},
  pages = {011601},
  numpages = {4},
  year = {2001},
  month = {Jun},
  publisher = {American Physical Society},
  doi = {10.1103/PhysRevLett.87.011601},
  url = {https://link.aps.org/doi/10.1103/PhysRevLett.87.011601}
}

@article{2001b_Felder,
  title = {Tachyonic instability and dynamics of spontaneous symmetry breaking},
  author = {Felder, Gary and Kofman, Lev and Linde, Andrei},
  journal = {Phys. Rev. D},
  volume = {64},
  issue = {12},
  pages = {123517},
  numpages = {19},
  year = {2001},
  month = {Nov},
  publisher = {American Physical Society},
  doi = {10.1103/PhysRevD.64.123517},
  url = {https://link.aps.org/doi/10.1103/PhysRevD.64.123517}
}

@article{2002_Copeland,
  title = {Dynamics of tachyonic preheating after hybrid inflation},
  author = {Copeland, E. J. and Pascoli, S. and Rajantie, A.},
  journal = {Phys. Rev. D},
  volume = {65},
  issue = {10},
  pages = {103517},
  numpages = {13},
  year = {2002},
  month = {May},
  publisher = {American Physical Society},
  doi = {10.1103/PhysRevD.65.103517},
  url = {https://link.aps.org/doi/10.1103/PhysRevD.65.103517}
}

@article{1986_Liu,
  author       = {I. Liu and I. Müller and T. Ruggeri},
  title        = {Relativistic thermodynamics of gases},
  journal      = {Annals of Physics},
  volume       = {169},
  number       = {1},
  pages        = {191--219},
  year         = {1986},
  doi          = {10.1016/0003-4916(86)90164-8},
}

@book{2013_RezzollaZanotti,
  author    = {Luciano Rezzolla and Olindo Zanotti},
  title     = {Relativistic Hydrodynamics},
  publisher = {Oxford University Press},
  year      = {2013},
  address   = {Oxford},
  isbn      = {978-0-19-852890-6},
  doi       = {10.1093/acprof:oso/9780198528906.001.0001}
}

\end{document}